\documentclass[journal]{new-aiaa}
\usepackage[utf8]{inputenc}
\usepackage{textcomp}

\usepackage{graphicx}
\usepackage{amsmath}
\usepackage[version=4]{mhchem}
\usepackage{siunitx}
\usepackage{longtable,tabularx}
\title{Asteroid Mining to Sustain a Mars Colony: \\A Logistics Point of View}

\author{Serena Suriano, Shamil Biktimirov\footnote{biktimirovshamil@gmail.com}, Dmitry Pritykin, Anton Ivanov}

\begin{document}

\maketitle

\begin{abstract}
Asteroid mining can become an enabling technology to establish a sustainable manned colony on Mars, which requires metallic materials more often than they are readily available in shipments from Earth. This paper describes a feasibility study of a supply chain that delivers metals extracted from metallic asteroids to Mars. The asteroids are selected to respect the $\Delta V$ limits imposed by up-to-date spacecraft. The study is conducted with reference to the state of the art in space transportation technologies and in-situ resource utilization. A possibility for mining on carbonaceous asteroids to produce the propellant required for return trips is also taken into account. Different supply chains are computed through a multi-objective optimization routine that considers the mission $\Delta V$, the mass of extracted metals and the mass of propellant produced on the asteroids. Schedules to visit the asteroids within reach are obtained and the total mass of the delivered material is evaluated for various mining rates. Finally, the  use of the metallic material to build habitats and rovers on the Martian soil through additive manufacturing is discussed.
 \end{abstract}

\section*{Nomenclature}


{\renewcommand\arraystretch{1.0}
\noindent\begin{longtable*}{@{}l @{\quad=\quad} l@{}}
$\Delta V$ & variation of velocity, km/s \\
$m$ & mass, kg \\
$\dot m$ & mass rate, kg/s \\
$I_{sp}$ & specific impulse, s \\
$\Delta t$ & waiting time, days \\
$\vec{V}$ & velocity in Cartesian coordinates, km/s \\
$\mu$ & standard gravitational parameter, km$^3$/s$^2$ \\
$r$ & radius in Cartesian coordinates, km \\
$a$ & semi-major axis, km \\
$e$ & eccentricity \\
$i$ & inclination, deg \\
$\omega$ & argument of perigee, deg \\
$\Omega$ & right ascension of the ascending node, deg \\
$MA$ & mean anomaly, deg \\
$T$ & UTC epoch \\

\end{longtable*}}

\section{Introduction}

Since the 1960s, dozens of unmanned space missions, including orbiters, landers, and rovers have been sent to Mars to collect data and answer questions about the red planet. Nowadays leading space agencies are discussing designs of a manned mission. The Mars Exploration Program Analysis Group (MEPAG) \citep{MEPAG} of the National Aeronautics and Space Administration (NASA) has established the scientific goals for Mars exploration for the coming years, one of such goals being preparation for human exploration. Systems engineering and design of a Mars Research Base is investigated by groups of researchers \citep{ruede2019, cichan2017}. Possibilities of either building an orbital station in Low Mars Orbit (LMO) as in M$^3$ project \citep{taraba200688}, or sending humans to the Mars surface as in Mars Direct proposal \citep{zubrin1992} are seriously studied. The sustainable stepwise approach in preparation for human presence on Mars is presented in \cite{price2015}, starting from a crewed mission to Phobos in the mid-2030's, proceeding towards short-term missions on Mars, and consummating with regular missions at a permanent Martian base in the 2040's.

Sending a crew to the Martian soil has several important advantages. The human presence would allow rapid visual identification of the geological context, determination of the similarities and differences between the rocks, identification of samples of exceptional scientific value and adaptation of the analysis procedures of the samples collected with the use of the analysis results for the subsequent collection \citep{levine2010}. It is also possible to avoid the issues for the remote control of Mars rovers from Earth such as delays between the command order and its execution.

Given the travel time from Earth, a colonial settlement should be nearly self-sufficient for extended periods, especially for primary needs such as air, water, energy, and food. The long-term colony should also be able to replicate industrial activities on Earth. Most of the consumable resources must be produced and recycled. In addition to basic necessities, the colony must be able to make industrial products for construction and repair. In particular, the ability to manufacture metals is fundamental to any technological civilization. By far the most accessible industrial metal present on Mars is iron  \citep{zubrin_settle}. Some of the metals used on Earth for alloys, such as Molybdenum, are present in a smaller percentage on Mars and for others as Boron the percentage is unknown. Since the plan is to settle Mars, once the colony is established, an additional source of metals is needed to satisfy the demand. Metals are of extreme importance making it possible to construct or repair objects and rovers. Longer mission duration increases the probability of component failures, therefore, the  spare parts required for confidence in system maintenance capability will occupy a significant portion of the overall system mass. In-situ manufacturing of spare parts has been proposed as a means to reduce this spares logistics mass \citep{chen202080, kading2015317}. Additive manufacturing techniques are also discussed in this regard \citep{kading2015317, Owens2015}.

We propose to consider asteroid mining as a supplementary source of resources required for the sustainable development of the Mars colony. In recent years, asteroid mining was much discussed, and asteroid mining has been proposed to complement Earth-based supplies of rare Earth metals and to supply resources in space, such as rare earth metals or water \citep{dorrington_supply, Asteroid_mining_NEA_2021}. 
Asteroid mining campaigns have been conceived and their logistics analyzed \citep{ho201651, Ishimatsu2016, Vergaaij2019435}. Availability of water-based propellants and metals for construction materials was discussed in \citep{Ross2001, Elvis201420}. However. the economical practicality of asteroid mining ventures appeared to be debatable when considering end-users on Earth \citep{hein2020104}. 

A preliminary study on supplying a base on Mars with resources ex\-trac\-ted from Near Earth Asteroids (NEA) has been proposed by \cite{Biktimirov2019}. We shall take its point further and examine the possibility of designing a supply chain for sending spaceships out from Mars to mine on Near Earth or Main Belt Asteroids (MBA) and bring the extracted metals to Mars. Mars crossing asteroids are also included into the selection scheme. In addition to metal extraction from the so-called metallic asteroids, our study highlights the need to also visit carbonaceous asteroids and produce extra propellant needed for return trips.

Meteorites exploration indicates that  small celestial bodies can be metal-rich \citep{Krispin2016}. General point of view to the origin of metal-rich asteroids is the formation of differentiated small planets with a metal-rich core, and following strip out of silicate crust/mantle by collisions with other small bodies \citep{Yang2007888}. However, new data collected in preparation of a mission to the largest M-asteroid (16) Psyche show a possibility of a complex mixed metal and silicate structure  \citep{Elkins-Tanton2020}.  Hence caution is needed while one estimates metal content astronomical spectral data only \citep{Neeley201437, Landsman2015186}. Metallic asteroids can be considered as a source of metallic iron-nickel alloys, ferrous sulphide minerals and olivine. Trace amounts of rare metals, such as Platinum group metals (PGM -ruthenium, rhodium, palladium, osmium, iridium, and platinum) can also be found. Carbonaceous asteroids can provide materials for in-situ propellant production (ISPP) due to their water content and potential for hydrocarbon production \citep{asteroid_book}. 

The aim of the research presented here is to design and optimize a single-product supply chain to sustain a Mars colony by mining asteroids for metals. The structure of the article is as follows. Section~2 describes the constituents of the proposed supply chain design such as spacecraft, chain-nodes and considered transfers. Section~3 outlines the asteroid selection routine which takes into account the required metallic asteroids and associates to them carbonaceous asteroids for multi-stage transfers. Section~4 analyzes different objective func\-tions for the optimization problem and outlines the schedule optimization pro\-ce\-dure. Section~5 presents the results for the optimized supply chain design and discusses how the materials brought to Mars can be put to use and lists key technologies required for the supply chain to operate. Finally, the last section concludes the paper.

\section{Problem statement}

\begin{figure}[hbt!]
\centering
\includegraphics[width=1\textwidth]{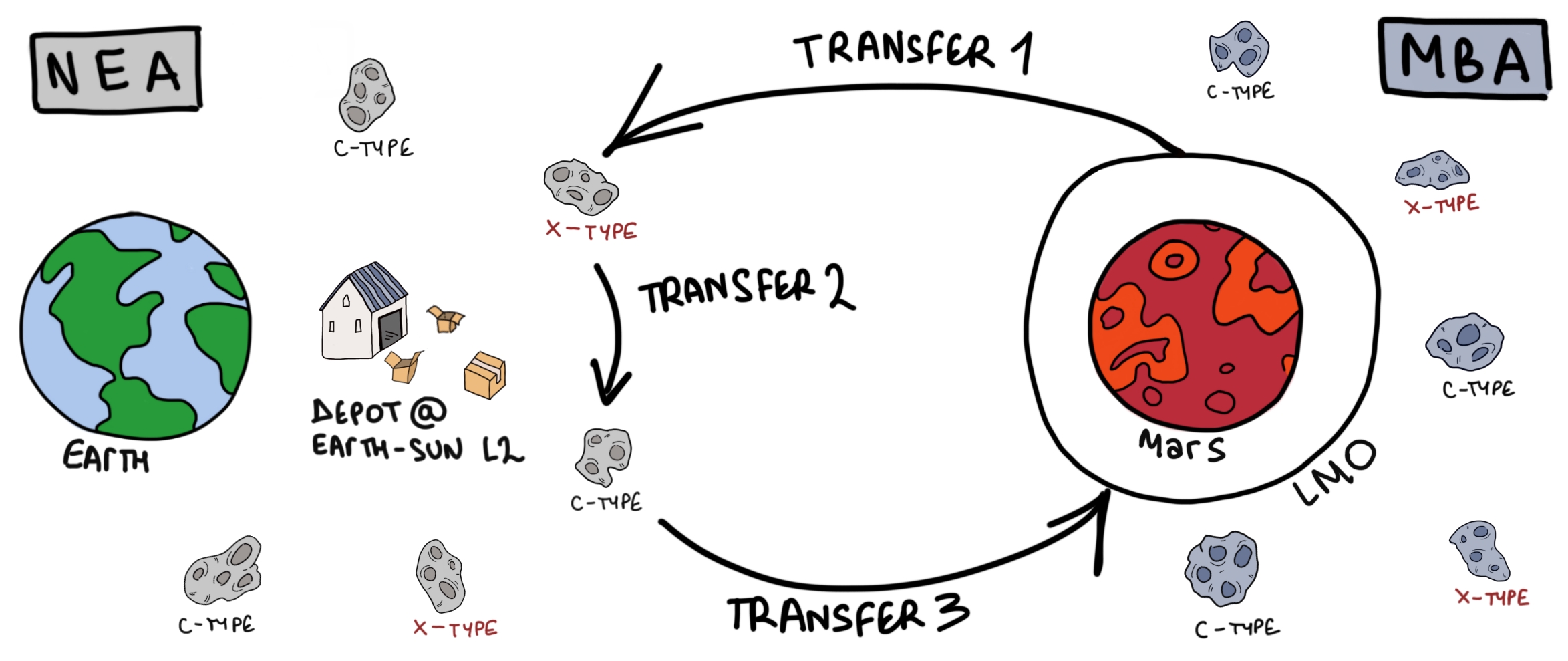}
\caption{General view of the concept of operations for supply of materials for a Martian colony from asteroids.}
\label{fig:logistics}
\end{figure}

The purpose of this article is to design and optimize a supply chain that delivers the metals extracted from metallic asteroids to Low Mars Orbit. This involves introduction of a fleet of cargo spacecraft, identification of the asteroids to travel to, and evaluation of a schedule for each cargo spacecraft, such that the delta-V is minimized and the mass of metals and propellant available are maximized. Such optimization problem formulation led us to use a multi-objective genetic algorithm \citep{multiobj_paper}. Application of this algorithm to our problem as well as several objective functions we introduce and analyze are described in the following sections. Overview of the concept of operations is shown in Fig.~\ref{fig:logistics}, which introduces the following key points: Low Mars Orbit, where materials are delivered from asteroids; Transfer 1 trajectory is for mining an X-type asteroid; Transfer 2 trajectory is to approach and mine C-type asteroid and Transfer 3 trajectory is go back to LMO; Depot at Earth-Sun L2 point along with its natural use is also a benchmark to compare the obtained routs to the direct delivery from Earth. 

Let us now introduce all the elements that comprise the tentative supply chain. It turns out that the key element in the logistics network is the cargo spacecraft, because it is the spacecraft's characteristics that drive the architecture of the network nodes (inclusion of carbonaceous asteroids along with the metallic ones), and impose additional requirements to the transfers during one trip (multiple hops instead of simple Mars-asteroid-Mars transfer). Thus, we shall start from describing the spacecraft.

\subsection{Spacecraft}
A cargo spacecraft with technical specifications similar to those of the Starship by SpaceX \cite{spacex} is considered.

\renewcommand{\tablename}{\textbf{Table}}
\captionsetup[table]{labelfont={bf},labelformat={default},labelsep=period,name={Table}}
\begin{table}[!htb]
\centering
\caption {Cargo spacecraft parameters~\cite{zubrin2019}} \label{tab:Starship} 
\begin{tabular}{ccccc}
\hline
Propellant mass & Dry mass & Payload mass & Specific impulse & $\Delta V$ \\ \hline
1100 ton        & 120 ton  & 115 ton      & 375 s            & 6.4 km/s                \\ \hline
\end{tabular}
\end{table}

As stated earlier, the choice of spacecraft sets limits to the transfer edges which have to  correspond to the available $\Delta V$. Our numerical experiments, described in detail in the section that deals with the asteroid selection, show that with the spacecraft characteristics as presented in the Table \ref{tab:Starship} there are no metallic asteroids from which the direct return trip to Mars can be guaranteed given the amount of fuel available on board. It is this fact that made us take into consideration carbonaceous asteroids on which the required propellant can be produced. Thus, for each mission, the idea is that of performing a first transfer from Mars to a metallic asteroid to mine metals on it, then move on to a carbonaceous asteroid to produce the propellant necessary for the return trip to reach Mars and deliver the cargo. 

It can be noted that in-situ propellant production system that can be deve\-loped on Mars \citep{kleinhenz2017} also imposes certain constraints on the system due to limited production rate. For this reason various research groups consider other options for thrust generation, such as solar sails \citep{VERGAAIJ20213045}, solar electric propulsion \citep{wooley201951} or nuclear propulsion \citep{nam2015678}. These approaches, although show distinct promise for space travel, still remain quite futuristic in the context of the problem at hand, and in this study we shall confine ourselves to the up-to-date technology as represented by the Starship \citep{spacex}.  

Note, that the mining equipment and the in-situ propellant production equipment are considered to be included in the structure of the cargo spacecraft for mass budget calculations. 

\subsection{Product}
The supply chain delivers metals as a single product. Any propellant is produced only with the aim to guarantee the feasibility of the return trip.

\subsection{Orbital Nodes}
The nodes of the supply chain are (Fig.~\ref{fig:logistics}):
\begin{itemize}
    \item Low Mars Orbit \\
    The LMO is circular and has altitude of \SI{500}{km}. We consider this orbit to be the starting point of all missions and the end point to which all cargo is delivered. 
    
    \item Metallic Asteroids\\
    The metallic asteroids are selected from NEAs and MBAs according to the criteria specified in the following sections. These asteroids supposedly allow extracting metals.
    
    \item Carbonaceous Asteroids\\
    The carbonaceous asteroids are also selected from NEAs and MBAs, se\-lec\-tion criteria are discussed and established in the following sections. These asteroids are assumed to allow producing propellant.
    
    \item Depot in Sun-Earth L2 point\\
    Having a depot at the Earth-Sun L2 point can make available metals at times when no launch windows to any of the selected asteroids are available. The L2 point is chosen since it is easily accessible from Earth thanks to their fixed relative position. It is assumed that cargo is delivered to this node from the Earth and can be picked up by a cargo spacecraft from Mars without any delay, which should be introduced, when time is needed to actually mine for metals on an asteroid. Using other libration points can be further considered, however we will confine our study to the Sun-Earth L2, because it comes closer to Mars than Earth-Moon L2 and easier to reach from Earth (to replenish the depot) than Sun-Mars L1. Including the Sun-Earth L2 point into the supply chain also allows us making comparisons of the Mars-based asteroid mining campaigns versus delivery of materials from Earth.
\end{itemize}

\subsection{Transfer Edges}

The transfers on which the multi-stage supply chain is based are (Fig.~\ref{fig:logistics}):

\begin{itemize}

    \item LMO - Metallic Asteroids\\
    The first transfer allows reaching a metallic asteroid from the LMO.
    
    \item Metallic Asteroids - Carbonaceous Asteroids\\
    The second transfer allows reaching a carbonaceous asteroid starting from the metallic asteroid.
    In this study, only the transfer from a metallic to a carbonaceous asteroid is considered. For this transfer the available $\Delta V$ is greater than the $\Delta V$ available for the reverse path. In fact, in the return trip the $\Delta V$ would be lower due to the high mass of propellant with which the tanks should be filled before flying to the metallic asteroid. These calculations are performed using the Tsiolkovsky equation.
    
    \item Carbonaceous Asteroids - LMO\\
    The third transfer reaches the LMO from the carbonaceous asteroid.
    
    \item LMO - Depot in L2\\
    This transfer is taken into consideration to guarantee the delivery of all those materials that cannot be extracted on asteroids and of those components that cannot be built on Mars.
    
\end{itemize}

\subsection{Problem Variables}
The optimal supply chain is achieved by solving a problem that takes into account the following variables:
\begin{itemize}
    \item Launch Date (LD) from LMO
    \item Launch Date from Metallic Asteroid
    \item Launch Date from Carbonaceous Asteroid
    \item Time of Flight (TOF) from LMO to Metallic Asteroid
    \item Time of Flight from Metallic Asteroid to Carbonaceous Asteroid
    \item Time of Flight from Carbonaceous Asteroid to LMO
    \item Stay time on Metallic Asteroid
    \item Stay Time on Carbonaceous Asteroid
    \item Metals mining rate
    \item Propellant Production rate
    \item Number of revolutions of the transfer orbit obtained by solving the Lam\-bert's problem
\end{itemize}

The stay time on asteroids consists of the time between the arrival on the asteroid and the next available launch window.  The launch dates and the time of flight together with the orbital parameters define the variation of velocity needed for the transfer. While the stay times and mining rates define the quantity of metals and propellant available.

\subsection{Procedure}
The procedure is summarized in the following steps.
\begin{itemize}
    \item Metallic asteroids selection;
    \item Association of one carbonaceous asteroid to each of the selected metallic asteroids;
    \item Multi-objective optimization of the supply schedule;
    \item Estimation of what can be produced depending on the available mass and the number of spacecraft used.
\end{itemize}

\section{Metal and Propellant Sources}

\subsection{Metallic Asteroids selection} \label{sect:ast_sel_X}
From the JPL small-body database \citep{jpl_small} the asteroids with the following features are selected:

\begin{table}[!htb]
\centering

\caption{Preliminary requirements for the metallic asteroids}\label{tab:preliminary_x}
\begin{tabular}{l l}
\hline
Orbit classes & \begin{tabular}[c]{@{}l@{}}Apollo, Amor, Mars-crossing Asteroid, \\ Inner MBA, MBA, Outer MBA\end{tabular} \\ \hline
Spectral type (SMASSII) & X, Xe, Xk, Xc \\ \hline
Minimum Diameter & 500 m
\end{tabular}
\end{table}

The orbit classes are selected such that the semi-major axis is $ a < \SI{4.6}{AU} $ and so the asteroids are not too far from Mars. Since the aim of the mission is providing Mars with metals, the spectral types selected correspond to metal-rich asteroids \citep{asteroid_book}.
A limit on the minimum diameter of \SI{500}{m} is imposed such that the asteroid is large enough for mining on it. 
For each asteroid, the $\Delta V$ map is computed \citep{vallado}.
The $\Delta V$ calculation depends on the Launch Date and the Time of Flight.
For each asteroid these two parameters are varied in order to obtain a $\Delta V$ matrix. 

Let us now explain the procedure in detail. The arrival date (AD) on the asteroid for the M-X trip (from Mars to metallic asteroid) is the sum of launch date (LD) and the time of flight (TOF):
\begin{equation}\label{eq:AD}
    AD = LD + TOF.
\end{equation}

The mean anomaly (MA) and, consequently, the true anomaly of the asteroid are updated according to the date considered 
\begin{equation}\label{eq:MA}
    MA = MA_{0} + \sqrt{\frac{\mu_{Sun}}{a_X^3}} (AD - T_{0}).
\end{equation}
\noindent where $MA_0$ is the initial mean anomaly, $\mu_{Sun}$ is the Sun gravity constant, $a_X$ is the semi-major axis of the X-asteroid.

Updating the orbital parameters as a function of the date and time of flight allows to define the position of the asteroid and Mars in Cartesian coordinates centered in the Sun.

At this point the Lambert's problem is solved taking into account several possible cases by varying 
\begin{itemize}
    \item orbit type: direct or retrograde transfer;
    \item number of revolutions : up to 5 revolutions;
    \item case: small-a  or large-a option.
\end{itemize}

In the context of the patched conic approximation, the escape manoeuvre $\Delta V_{esc}$ around Mars is calculated as follows:

\begin{equation}\label{eq:v_inf}
V_{\infty} = ||\Vec{V}_{S/C} - \vec{V}_{M}||,
\end{equation}

\begin{equation}\label{eq:escape}
\Delta V_{esc} = \sqrt{{V_{\infty}}^2 +  \frac{2\mu_{M}}{r_{LMO}}} - \sqrt{\frac{\mu_{M}}{r_{LMO}}},
\end{equation}

\noindent where $V_{\infty}$ is the spacecraft hyperbolic excess velocity, $V_{S/C}$ is the spacecraft velocity at heliocentric transfer trajectory, $V_M$ is heliocentric velocity of Mars, $\mu_{M}$ is the gravitational parameter of Mars, $r_{LMO}$ - is the radius of Low Mars Orbit.

Since the gravitational contribution of the asteroid is negligible the capture manoeuvre $\Delta V_{arr}$ at an X-type asteroid is approximated as follows:  

\begin{equation}\label{eq:arr}
    \Delta V_{arr} = || \Vec{V}_{S/C} - \vec{V}_{X}||,
\end{equation}

\noindent where $V_X$ is the heliocentric velocity of an X-type asteroid.

The orbital transfers are modelled as a series of impulsive changes of velocity.
Finally, the lowest $\Delta V_{LMO-X}$ \eqref{eq:dV_total} is selected among all the $\Delta V_{LMO-X}$ obtained for all the cases examined while solving the Lambert's problem. 

\begin{equation}\label{eq:dV_total}
    \Delta V_{LMO-X} = \Delta V_{esc} + \Delta V_{arr}.
\end{equation}

\begin{figure}[!htb]
 \begin{center}
  \includegraphics[width=1\textwidth]{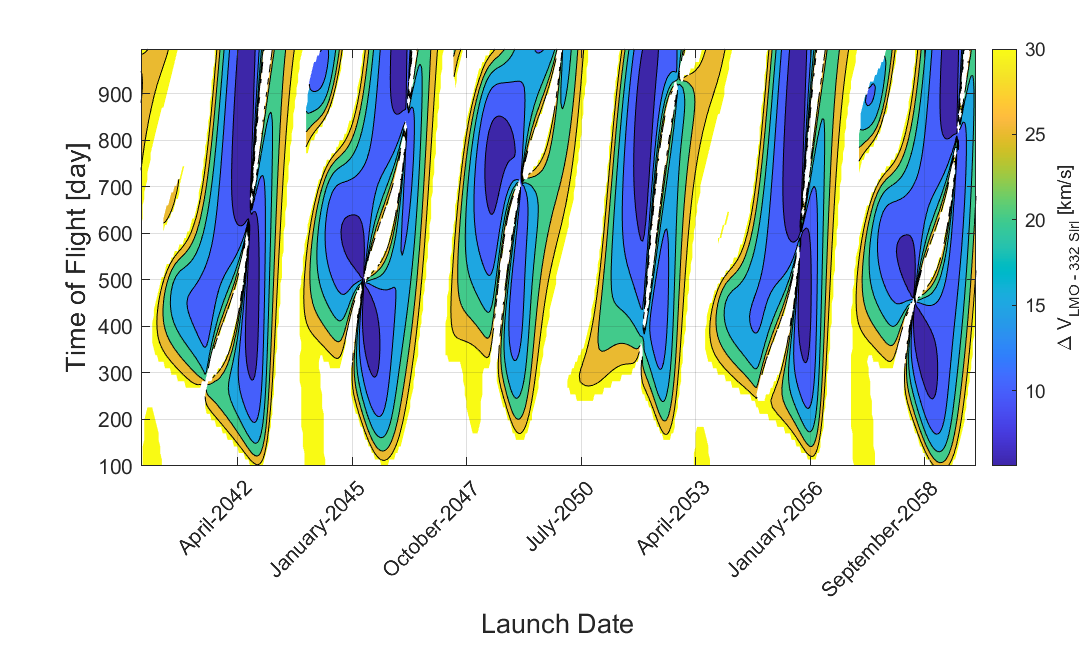}
  \caption{$\Delta V$ map of the transfer from LMO to 332 Siri.}
  \label{fig:map_MX}
   \end{center}
\end{figure}

\begin{figure}[!htb]
 \begin{center}
  \includegraphics[width=1\textwidth]{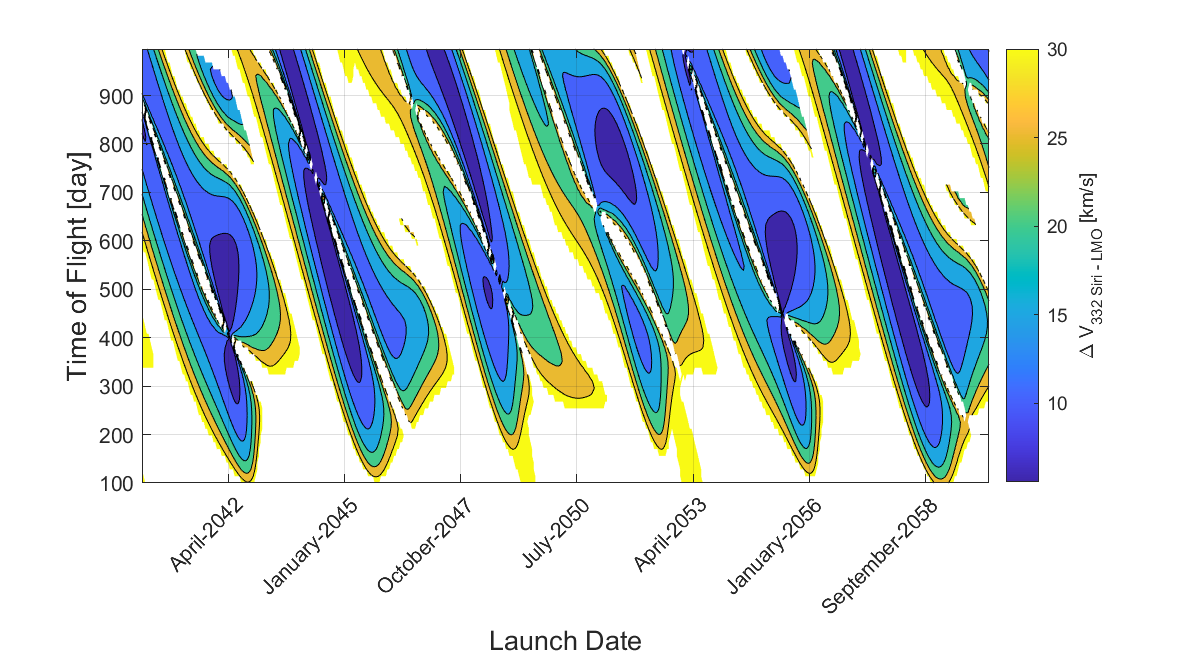}
  \caption{$\Delta V$ map of the transfer from 332 Siri to LMO.}
  \label{fig:map_XM}
   \end{center}
\end{figure}

This is done for several LD and TOF. The $\Delta V$ map is evaluated for Times of Flight ranging from 100 to 1000 days and Launch Dates in a 20 years interval \footnote{About the same timescale needed for the assembly of the International Space Station, whose mass is approximately equal to the mass we want to deliver at LMO} starting from January $1^{st}$ 2040. The $\Delta V$ maps for the asteroid ``332 Siri'' ($a=2.774AU, e=0.0891, i=2.875$) is reported in Fig.~\ref{fig:map_MX} - \ref{fig:map_XM}.

Note that the correctness of the $\Delta V$ map calculation algorithm has been verified with the JPL Small-Body Mission-Design Tool \citep{jpl_mission} in which the departure from Earth is considered. 

Initially, the possibility of a round trip to reach a metallic asteroid is ex\-am\-ined. Given the limits of propellant available for the transfer vehicle (see Table~\ref{tab:Starship}), the maximum $\Delta V$ for the two transfers is equal to \SI{6.4}{km/s}. 

\begin{figure}[!htb]
 \begin{center}
  \includegraphics[width=0.8\textwidth]{ 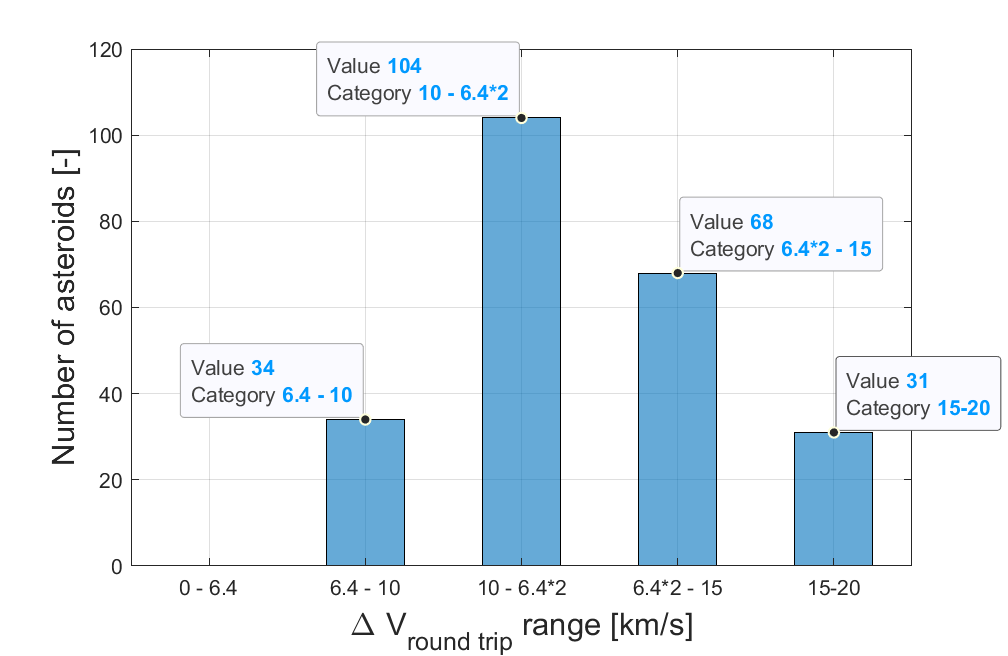}
  \caption{Metallic asteroids sorted by $\Delta V_{round trip}$ from LMO.}
  \label{fig:bar_x}
   \end{center}
\end{figure}

Fig.~\ref{fig:bar_x} characterizes pre-selected metallic asteroids that meet the preliminary requirements (Table \ref{tab:preliminary_x}).
For each of these asteroids, the $\Delta V$ of the round trip is evaluated. It is shown that there are no asteroids reachable by considering $\Delta V_{round \ trip} = \SI{6.4}{km/s}$. Most of them have a $\Delta V_{round \ trip}$ between 10 and \SI{12.8}{km/s} (that is the double of the maximum capability). For this reason, refueling is required to guarantee the return trip to LMO.

\begin{figure}[!htb]
 \begin{center}
  \includegraphics[width=1\textwidth]{ 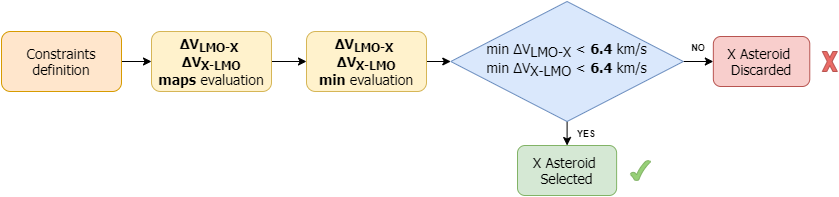}
  \caption{Flowchart of the Metallic Asteroids selection.}
  \label{fig:flowX}
   \end{center}
\end{figure}

The idea is, therefore, to reach a metallic asteroid with a limit of $\Delta V$ equal to 6.4 km/s for the transfer from Mars to the metal-type asteroid and then produce propellant on a carbonaceous asteroid. This leads to the last constraint, that is the propellant capability of the transfer vehicle. The  selection consists in choosing the metallic asteroids reachable with a maximum $\Delta V$ of \SI{6.4}{km/s}, from which it is possible to return to LMO with \SI{6.4}{km/s}. They are 122. A flowchart to explain the procedure is presented in Fig.~\ref{fig:flowX}. 

Notice that in Fig.~\ref{fig:bar_x} the number of asteroids with a total variation of velocity lower than \SI{12.8}{km/s} is 138, whereas there are only 122 selected asteroids. This is because, among the 138, there are asteroids with $\Delta V_{round \ trip} < \SI{12.8}{km/s}$ and $\Delta V_{single \ trip}>\SI{6.4}{km/s}$ that are discarded.

\subsection{Carbonaceous Asteroids selection}
\label{sect:ast_sel_C}
Given the capability of the cargo spacecraft in terms of $\Delta V$, a way to ensure the refueling is studied. For this reason, to each metallic asteroid a carbonaceous asteroid is associated, with the following procedure.

\begin{figure}[!htb]
\centering
 \begin{center}
  \includegraphics[width=1\textwidth]{ 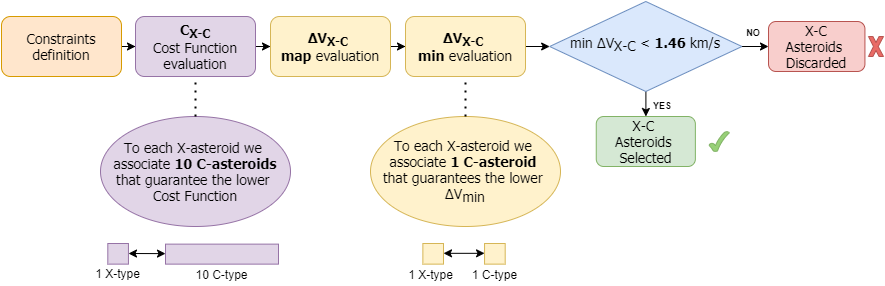}
  \caption{Flowchart of the carbonaceous asteroids selection and association with the metallic asteroids.}
  \label{fig:flowC}
   \end{center}
\end{figure}

From the JPL small-body database \citep{jpl_small}, the carbonaceous asteroids with the following features are pre-selected. 

\begin{table}[h!]
\centering
\label{tab:c}
\caption{Preliminary Requirements for carbonaceous asteroids.}
\begin{tabular}{l l}
\hline
Orbit classes & \begin{tabular}[c]{@{}l@{}}Apollo, Amor, Mars-crossing Asteroid, \\ Inner MBA, MBA, Outer MBA\end{tabular} \\ \hline
Spectral type (SMASSII) & C, B, Cg, Cgh, Ch, Cb \\ \hline
Minimum Diameter & 500 m
\end{tabular}
\end{table}

The orbit classes are selected such that the semi-major axis is $ a < \SI{4.6}{AU} $ and so the asteroids are not too far from Mars. The spectral types selected correspond to water-rich asteroids \citep{asteroid_book}.

For each metallic asteroid, a set of 10 possible metallic-carbonaceous pairs is created such as to minimize a cost function that takes into account the variation of velocity of the Hohmann transfer \citep{vallado} between the two orbits and the variation of the orbital parameters normalized by the orbital parameters of the metallic asteroid. This is given by 

\begin{equation}\label{eq:v_inf_}
C_{X-C} = \Delta V_{H}+
\frac{\Delta a}{a_{X}} + \frac{\Delta e}{e_{X}} + \frac{\Delta i}{i_{X}} + \frac{\Delta \Omega}{\Omega_{X}} + \frac{\Delta \omega}{\omega_{X}},
\end{equation}
\noindent where the subscripts $X$ and $C$ stand for metallic and carbonaceous asteroids, respectively.

\begin{figure}[!htb]
 \begin{center}
  \includegraphics[width=0.8\textwidth]{ 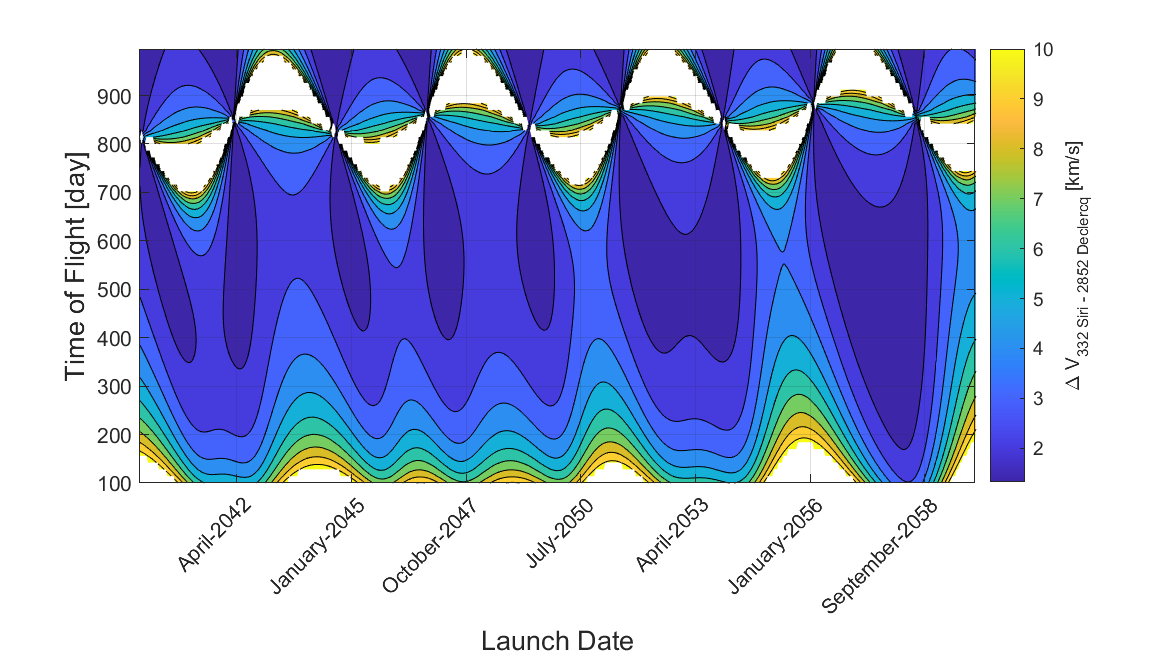}
  \caption{$\Delta V$ map of the transfer from a metallic asteroid to a carbonaceous asteroids.}
  \label{fig:map_XC}
   \end{center}
\end{figure}

Then for all the initial pairs, that are $122\times10$ the $\Delta V$ maps are computed within the same time ranges and assumption stated before. Fig.~\ref{fig:map_XC} shows the $\Delta V$ map for the transfer from one metallic asteroid to its associated carbonaceous asteroid. 

Only one pair for each metallic asteroid is taken into account. It is the one that guarantees the lowest $\Delta V$ from the metallic asteroid to the carbonaceous one. 
The propellant for this transfer can be carried as a payload in the trip from Mars to the metallic asteroid, so $m_{prop}=\SI{115}{ton}$. Given also $m_{s}=\SI{120}{ton}$ and  $m_{pay}=\SI{115}{ton}$, from Table \ref{tab:Starship}, the maximum variation of velocity is computed by applying the Tsiolkovsky equation: 

\begin{equation}\label{eq:Tsiolkovsky}
\Delta V_{X-C}^{max} = I_{sp}\, g_{0}\, \ln\biggl(\frac{m_s+m_{pay}+m_{prop}}{m_s+m_{prop}}  \biggr).
\end{equation}

The maximum $\Delta V$ available for the transfer from the metallic asteroid to the carbonaceous asteroid is \SI{1.46}{km/s}. This further limit leads to the final selection of possible pairs of asteroids on which to extract metals and create propellant to return to Mars. Applying all the constraints we obtain 22 pairs of asteroids: 22 metallic asteroids and 19 carbonaceous asteroids. The couples are listed in Appendix~A. 

\subsection{Depot}

The $\Delta V_{LMO-L2}$ is computed using the following equation for the Depot semi-major axis \citep{L2}: 
\begin{equation}\label{eq:L2}
a_{L2} = \SI{1.01}{AU}.
\end{equation}

The $\Delta V$ map is evaluated with the same timescale as previously used. The transfers are: LMO $\rightarrow$ L2 and L2 $\rightarrow$ LMO.
The launch windows are selected such that $\Delta V_{LMO-L2} < \SI{6.4}{km/s}$ and $\Delta V_{L2-LMO} < \SI{6.4}{km/s}$.

\section{Schedule Optimization}
 
 This section describes the procedure to compute the optimized schedule in these cases:
 
 \begin{enumerate}
     \item Metallic asteroids are visited\\
     A) minimum mining rate\\
     B) maximum mining rate
     \item Pairs metallic-carbonaceous asteroids are visited\\
     A) minimum mining rate\\
     B) maximum mining rate
     
     \item Metallic asteroids and the Depot are visited\\
     A) minimum mining rate\\
     B) maximum mining rate
     
     \item Pairs metallic-carbonaceous asteroids and the Depot are visited\\
     A) minimum mining rate\\
     B) maximum mining rate
     
 \end{enumerate}

\subsection{Metallic Asteroids in the schedule}
In this case the carbonaceous asteroids are not considered to produce pro\-pel\-lant. It is not taken into account that the propellant for the return trip is unavailable. For each one of the 122 metallic asteroid that satisfy all the requirements, the launch windows that guarantee the departure-after-arrival constraint are evaluated. For each metallic asteroid the transfers are: LMO $\rightarrow$ metallic asteroid $\rightarrow$ LMO.
 
The aim of optimizing the supply chain is to guarantee the lowest possible cost in terms of propellant and the highest possible mass of extracted metal \citep{dorrington_supply}. To achieve this, two objective functions are defined: 
\begin{equation}\label{eq:f1x}
 \Delta V_{schedule}= \sum_{k=1}^N \Delta V_{LMO-X}^{(k)} + \Delta V_{X-LMO}^{(k)},
\end{equation}

\begin{equation}\label{eq:f2}
m_{metals} = \dot{m}_{mining} \; \Delta t_{X}.
\end{equation}

Eq.~\eqref{eq:f1x} takes into account the total $\Delta V$ of the schedule and $N$ stands for the number of metallic asteroid visited in one schedule. Eq.~\eqref{eq:f2} refers to the product between the mining rate and the stay time on the metallic asteroid. The first objective function (Eq.~\eqref{eq:f1x}) is minimized and the second one (Eq.~\eqref{eq:f2}) is maximized.

Note that the limitation on the mass payload specified in Table~\ref{tab:Starship} is already taken into account in the definition of the second objective function by imposing a maximum mined mass of \SI{115}{ton}. 

The multi-objective genetic algorithm (GA) implemented in \cite{multiobj_matlab} is ap\-plied with the settings specifically shown in Appendix B - C. The algorithm uses a controlled, elitist genetic algorithm (a variant of NSGA-II as described by \cite{multiobj_paper}). An elitist GA always favors individuals with better fitness value. A controlled elitist GA also favors individuals that can help increase the diversity of the population even if they have a lower fitness value. 
Problems which appear to be particularly appropriate for solution by GA include timetabling and scheduling problems.

In summary, a GA consist of these steps:
\begin{enumerate}
    \item evaluation of different starting solutions (as if they were different humans)
    \item recombination of the starting solutions (similar to biological reproduction)
    \item introduction of elements of disorder (similar to random genetic mutations)
    \item evaluation of new solutions (new people) by choosing the best ones (en\-vi\-ron\-men\-tal selection) in an attempt to converge towards optimal solutions
\end{enumerate}
Each of these phases of recombination and selection can be called generation.

In a multi-objective GA the algorithm works by starting from a certain number of possible solutions (the population). It tries to identify, through different iterations, a certain number of optimal solutions, which corresponds to the Pareto front. The difference with the one-objective GA lies in the fact that there are more fitness functions to evaluate.

Since the objective functions strongly depend on the mining rate, both the minimum and the maximum mining rate estimated in \cite{dorrington_rate} are considered. The mining rate goes from 100 to \SI{800}{kg/day}.

\subsection{Metallic-Carbonaceous Asteroids in the schedule} 
To guarantee the production of propellant for the return trip at LMO, the carbonaceous asteroids are visited in the schedule. 
For each pair of asteroids selected and for the three transfers (LMO $\rightarrow$ metallic asteroid $\rightarrow$ carbonaceous asteroid $\rightarrow$ LMO), the launch windows that guarantee the departure-after-arrival constraint are evaluated. 
The multi-objective optimization allows to minimize  the cost in terms of variation of velocity  
\begin{equation}\label{eq:f1xc}
 \Delta V_{schedule}= \sum_{k=1}^N \Delta V_{LMO-X}^{(k)} + \Delta V_{X-C}^{(k)} + \Delta V_{C-LMO}^{(k)},
\end{equation}

maximize the mass of extracted metal 
\begin{equation}\label{eq:f2xc}
m_{metals} = \dot{m}_{mining} \; \Delta t_{X}
\end{equation}

or maximize the the mass of propellant produced 
\begin{equation}\label{eq:f3}
m_{propellant} = \dot{m}_{ISPP} \; \Delta t_{C}.
\end{equation}

The values for the problem setup and the options set for the multi-objective optimization are shown in Appendix~B and Appendix~C, respectively. Note that the constraint for the mass payload is already taken into account in the definition of the second objective function by imposing a maximum mined mass of \SI{115}{ton}. 

Both the minimum mining rate of \SI{100}{kg/day} and the maximum mining rate of \SI{800}{kg/day} \citep{dorrington_rate} are considered. As regards the propellant production rate, the estimated value for a manned mission to Mars is taken into consideration, which is equal to \SI{2}{kg/day} \citep{zubrin2019}.

A further constraint can be added to guarantee a worthwhile schedule. The schedules with a small amount of mass mined could be neglected in order to guarantee an effective mission. In particular for each return transfer on LMO, a minimum mass of material to be delivered is imposed. This lower bounds are estimated with the aim of being feasible depending on the mining rate. The lower limits imposed on the mined mass returned on LMO are explained in Table \ref{tab:limit_10_80} and they are imposed both in the ``Metallic Asteroids'' and ``Metallic-Carbonaceous Asteroids'' cases.

\begin{table}[!htb] 
\centering
\caption{Lower limits of mined mass}
\label{tab:limit_10_80}
\begin{tabular}{cc}
\hline
Mining Rate Cases & Lower Limit of Mined Mass \\ \hline
\begin{tabular}[c]{@{}c@{}}Minimum Mining Rate\\ (100 kg/day)\cite{dorrington_rate}\end{tabular} & 10 ton \\ \hline
\begin{tabular}[c]{@{}c@{}}Maximum Mining Rate\\ (800 kg/day)\cite{dorrington_rate}\end{tabular} & 80 ton \\ \hline
\end{tabular}
\end{table}

\subsection{Asteroids and Depot in the schedule}

When considering the Sun-Earth L2 point as a source thanks to the depot, the presence of all the needed sources is assumed. The spacecraft is filled with metals up to its full capacity and the tanks are completely filled with fuel. 

\begin{equation}\label{eq:depot1}
m_{metals @ depot} = \SI{115}{ton},
\end{equation}

\begin{equation}\label{eq:depot2}
m_{propellant @ depot} = \SI{1100}{ton}.
\end{equation}

The optimization is conducted by adding the depot as a new source and considering the different cases in terms of mining rates. The optimization setup and options are shown in Appendix~B and Appendix~C, respectively.

\section{Results}

The results of the optimization conducted for the several cases are shown in detail in this section and summarized in Appendix D and Appendix E.

\begin{figure}[!htb]
 \begin{center}
  \includegraphics[width=0.8\textwidth]{ 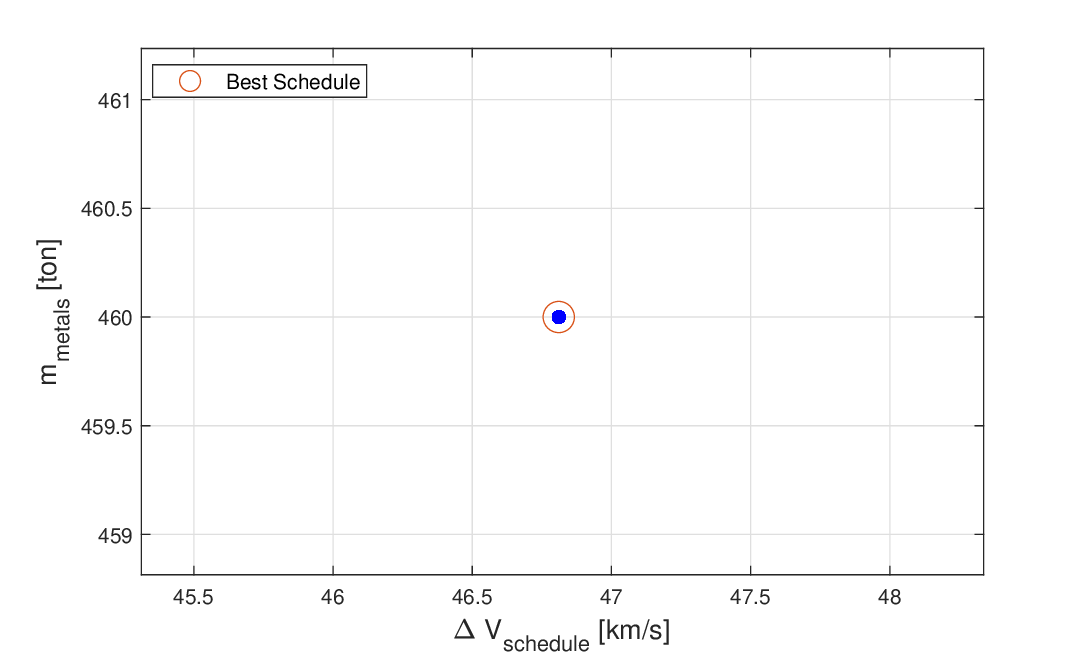}
  \caption{Pareto front with maximum mining rate if metallic asteroids are visited with the lower bound of \SI{80}{ton} on the mass delivered for each trip.}
  \label{fig:p_800_80_xxxx}
   \end{center}
\end{figure}

\begin{figure}[!htb]
 \begin{center}
  \includegraphics[width=0.8\textwidth]{ 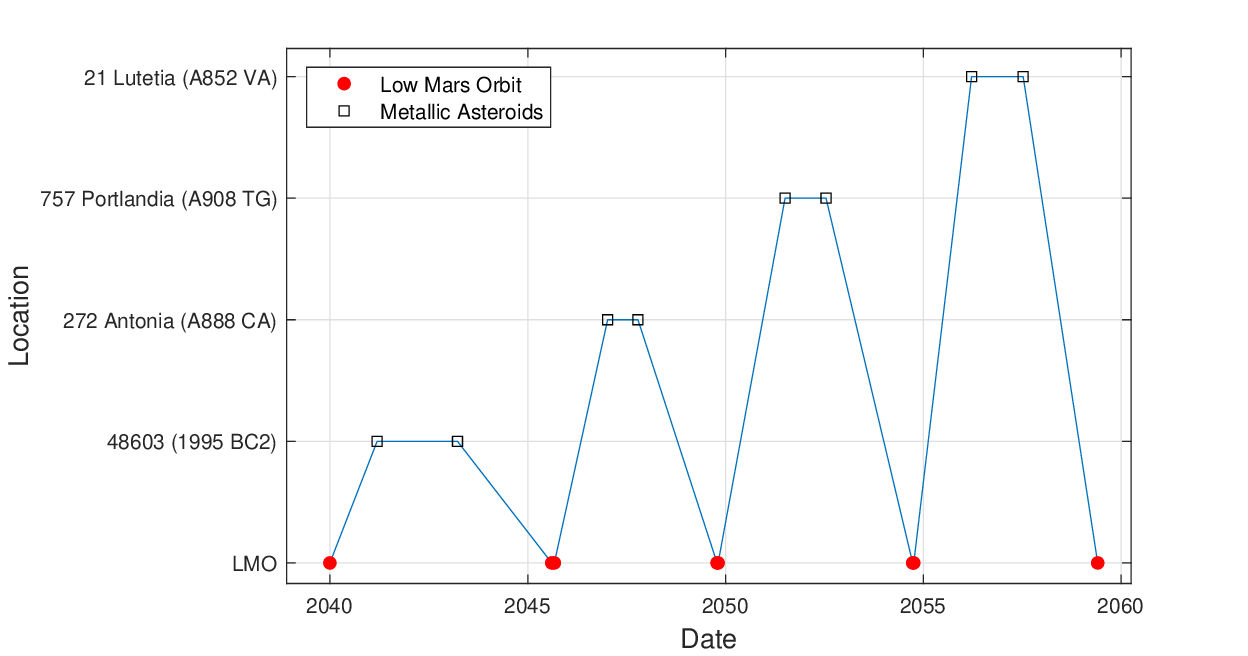}
  \caption{Best schedule with maximum mining rate if metallic asteroids are visited with the lower bound of \SI{80}{ton} on the mass delivered for each trip.}
  \label{fig:s_800_80_xxxx}
   \end{center}
\end{figure}

\begin{figure}[!htb]
\centering
 \begin{center}
  \includegraphics[width=0.8\textwidth]{ 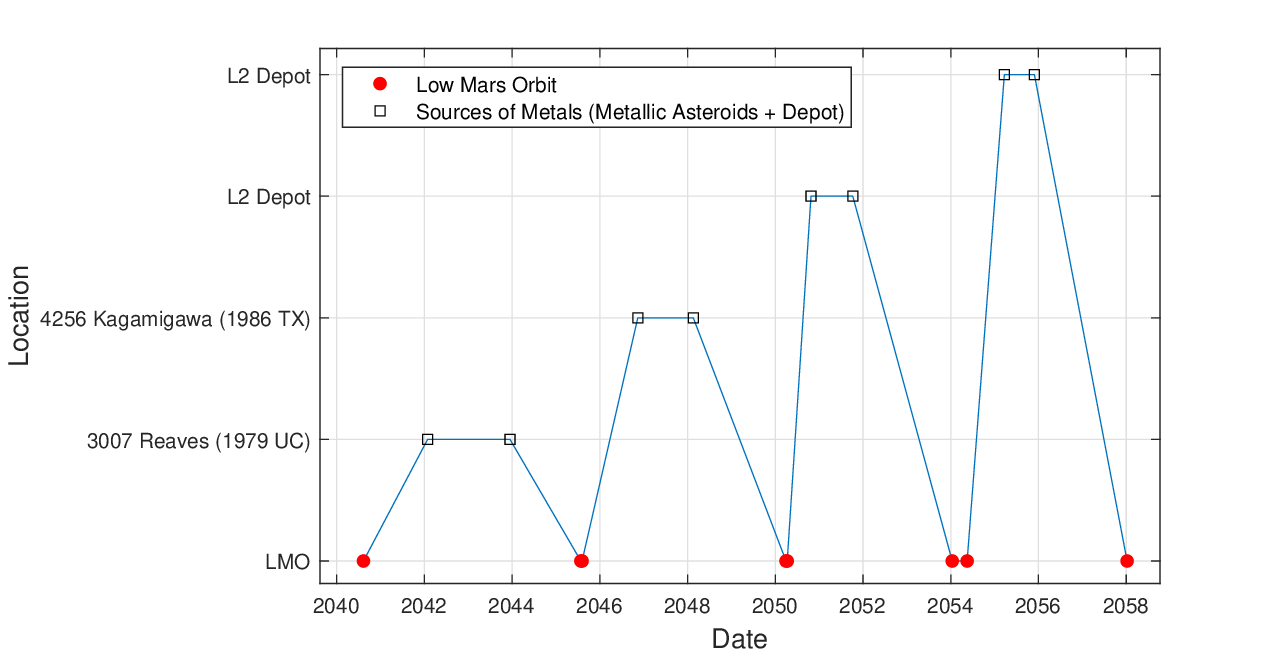}
  \caption{Best schedule with minimum mining rate if depot and metallic asteroids are visited with no lower bound on the mass delivered.}
  \label{fig:xL2_100_0_bs}
   \end{center}
\end{figure}

\begin{figure}[!htb]
\centering
 \begin{center}
  \includegraphics[width=0.8\textwidth]{ 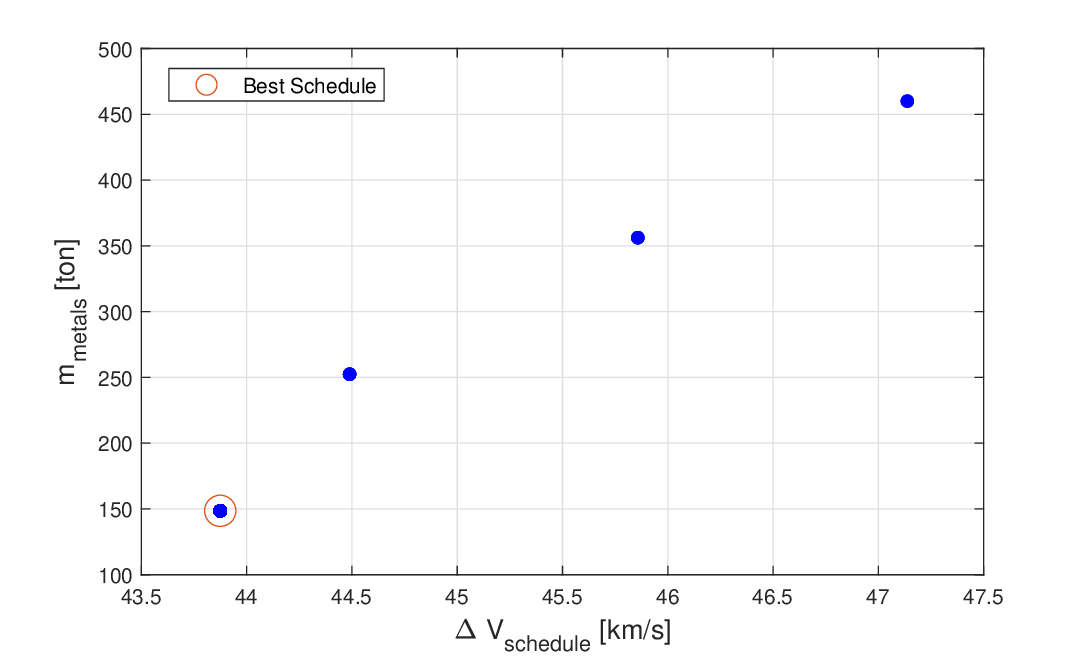}
  \caption{Pareto front with maximum mining rate if depot and metallic asteroids are visited with no lower bound on the mass delivered.}
  \label{fig:p_xL2_800_0}
   \end{center}
\end{figure}

\begin{figure}[!htb]
\centering
 \begin{center}
  \includegraphics[width=0.8\textwidth]{ 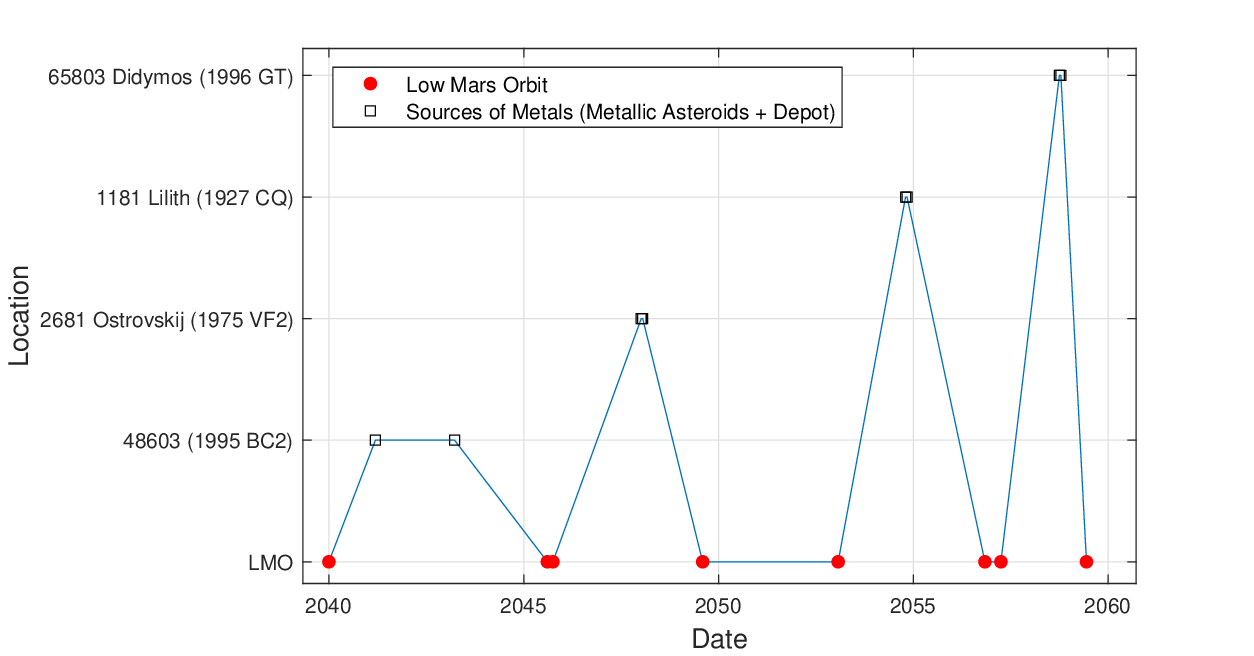}
  \caption{Best schedule with maximum mining rate if depot and metallic asteroids are visited with no lower bound on the mass delivered.}
  \label{fig:xL2_800_0}
   \end{center}
\end{figure}

\begin{figure}[!htb]
\centering
 \begin{center}
  \includegraphics[width=0.8\textwidth]{ 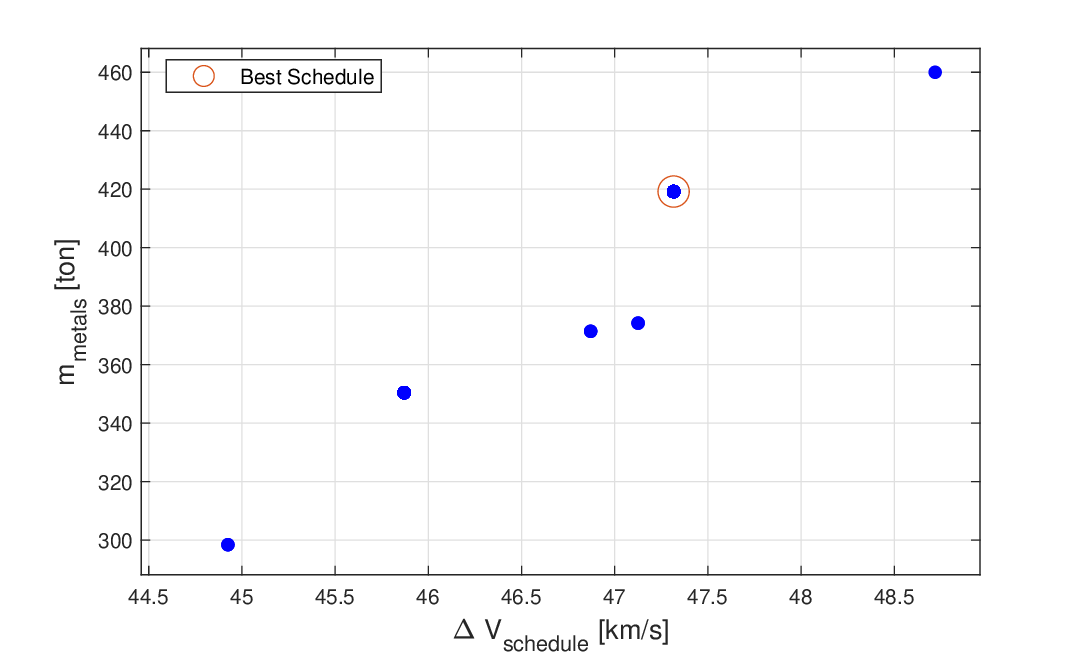}
  \caption{Pareto front with minimum mining rate if depot and metallic asteroids are visited with the lower bound of \SI{10}{ton} on the mass delivered for each trip.}
  \label{fig:xL2_100_10_pf}
   \end{center}
\end{figure}

\begin{figure}[!htb]
\centering
 \begin{center}
  \includegraphics[width=0.8\textwidth]{ 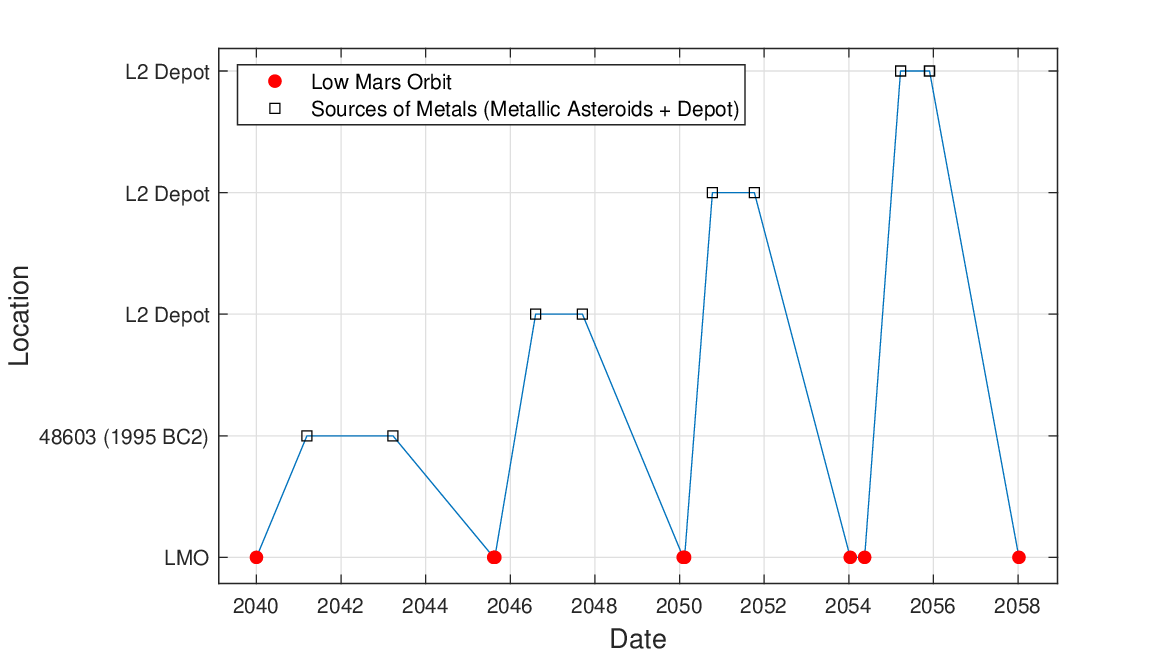}
  \caption{Best schedule with minimum mining rate if depot and metallic asteroids are visited with the lower bound of \SI{10}{ton} on the mass delivered for each trip.}
  \label{fig:xL2_100_10_bs}
   \end{center}
\end{figure}

\begin{figure}[!htb]
\centering
 \begin{center}
  \includegraphics[width=0.8\textwidth]{ 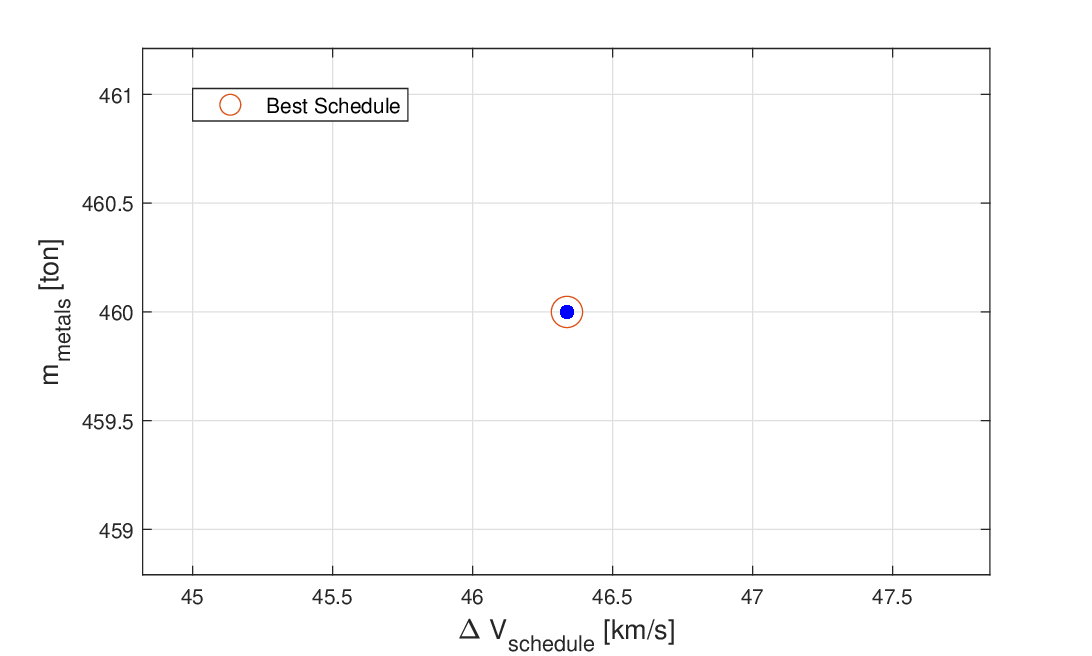}
  \caption{Best schedule with minimum mining rate if depot and metallic asteroids are visited with the lower bound of \SI{80}{ton} on the mass delivered for each trip.}
  \label{fig:p_xL2_800_80}
   \end{center}
\end{figure}

\begin{figure}[!htb]
\centering
 \begin{center}
  \includegraphics[width=0.8\textwidth]{ 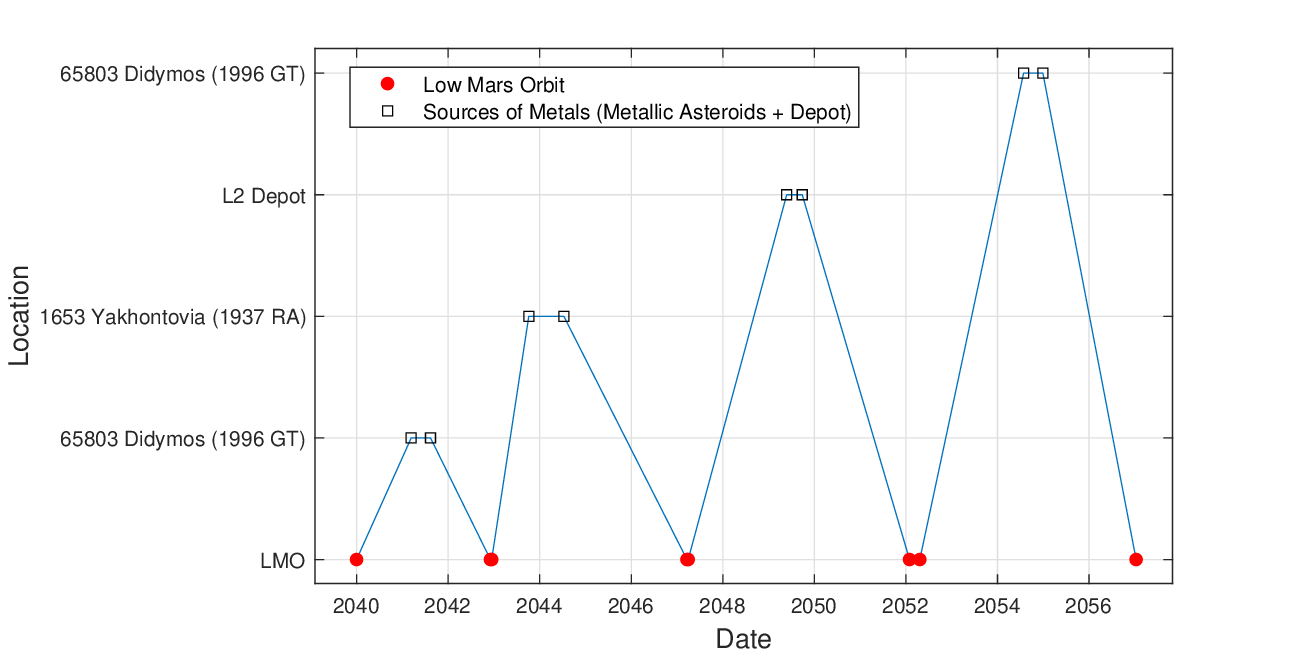}
  \caption{Best schedule with minimum mining rate if depot and metallic asteroids are visited with the lower bound of \SI{80}{ton} on the mass delivered for each trip.}
  \label{fig:xL2_800_80}
   \end{center}
\end{figure}

The results below are presented in three parts:
\begin{itemize}
    \item schedules including only metallic asteroids, Fig~\ref{fig:pareto100XXXX}-\ref{fig:s_800_80_xxxx}
    \item schedules including pairs of metallic and carbonaceous asteroids, where each pair is visited during a single trip, Fig~\ref{fig:p_100_0_xcxc}-\ref{fig:s_800_80_xcxc}
    \item schedules including the L2 depot along with the asteroids, Fig~\ref{fig:xL2_100_0_bs}-\ref{fig:xL2_800_80}
\end{itemize}

For each part the optimization are conducted for the following conditions, as given by Table~\ref{tab:limit_10_80}:
\begin{itemize}
    \item minimum mining rate
    \item maximum mining rate
    \item lower bound of metals mass collected from an asteroid with minimum mining rate - 10~ton    
    \item lower bound of metals mass collected from an asteroid with maximum mining rate - 80~ton
\end{itemize}

For each set of parameters we provide the plot of Pareto front and a plot with the best schedule.

\subsection{Metallic Asteroids in the schedule}

The case in which only metallic asteroids are visited is computed. This is not feasible by using the above mentioned cargo spacecraft, but it would be thanks to a spacecraft that could guarantee, without refueling, the double of the $\Delta V$, that is \SI{12.8}{km/s}. In this way the amount of propellant is sufficient for the round trip. 

\begin{figure}[!htb]
 \begin{center}
  \includegraphics[width=0.8\textwidth]{ 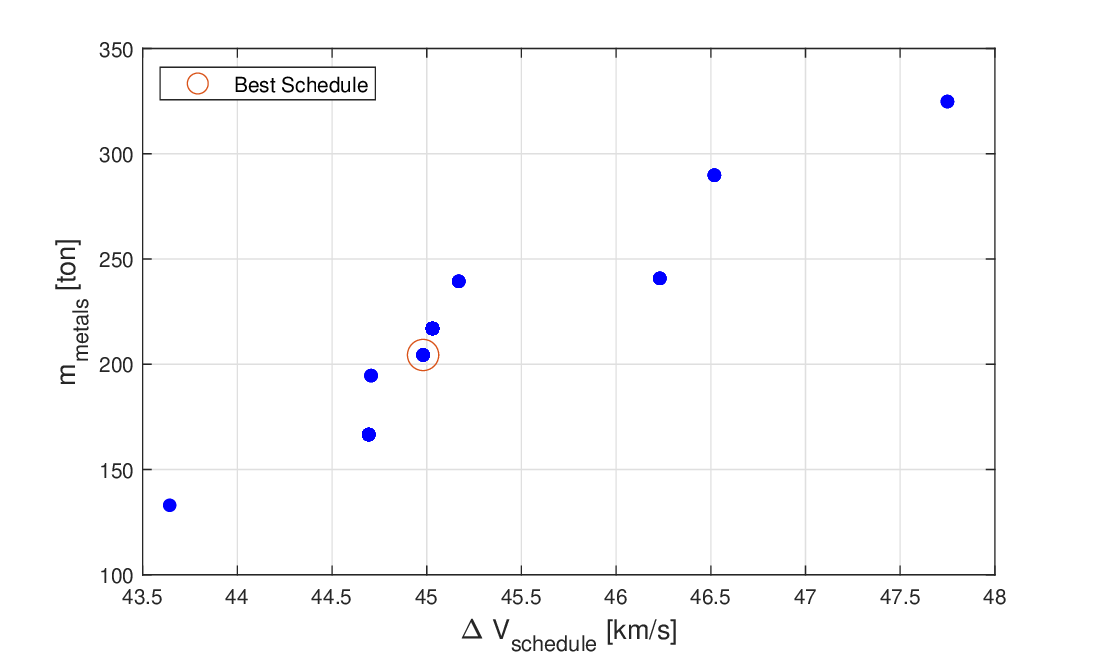}
  \caption{Pareto front with minimum mining rate if 4 metallic asteroids are visited.}
  \label{fig:pareto100XXXX}
   \end{center}
\end{figure}

\begin{figure}[!htb]
 \begin{center}
  \includegraphics[width=0.8\textwidth]{ 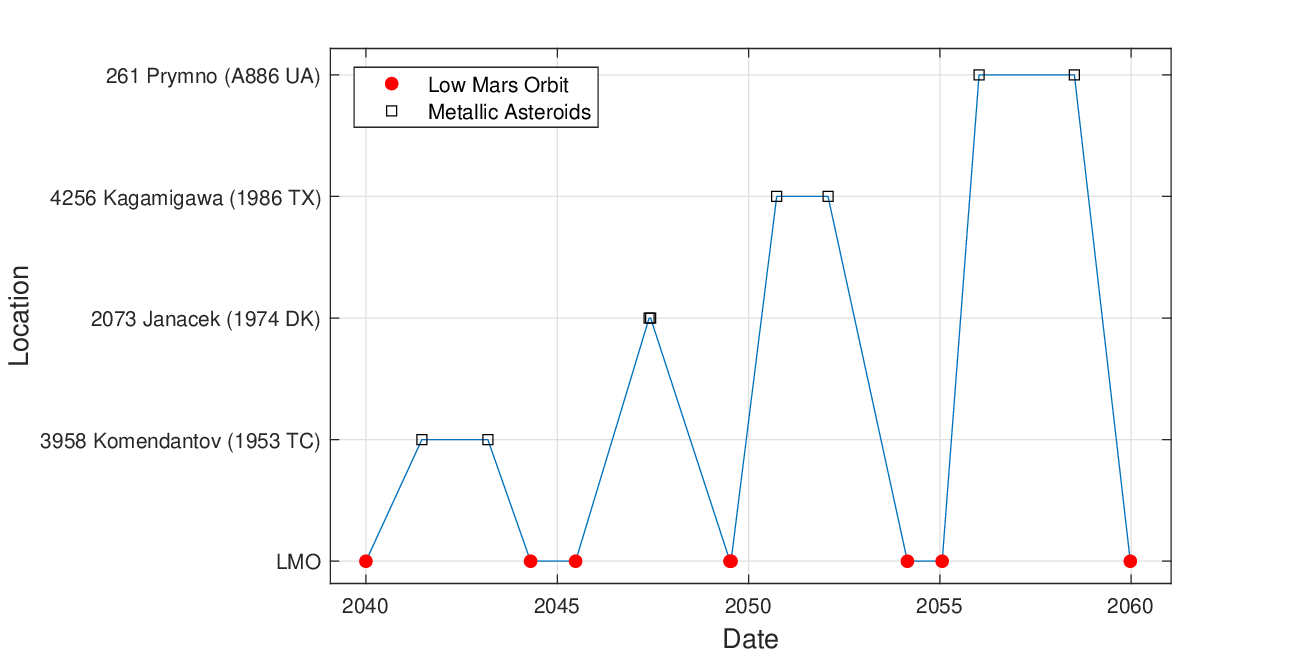}
  \caption{Best schedule with minimum mining rate if only metallic asteroids are visited.}
  \label{fig:schedule100XXXX}
   \end{center}
\end{figure}

The schedule to minimize $\Delta V$ and maximizes the mass of metals extracted, allows to mine on 4 metallic asteroids in 20 years with one transfer vehicle. The Pareto front of the multi-objective optimization conducted with the minimum mining rate of \SI{100}{kg/day} \cite{dorrington_rate} is shown in Fig.~\ref{fig:pareto100XXXX}. The schedule corresponding to the best solution given by the Pareto front is presented in Fig.~\ref{fig:schedule100XXXX}.

\begin{figure}[!htb]
 \begin{center}
  \includegraphics[width=0.8\textwidth]{ 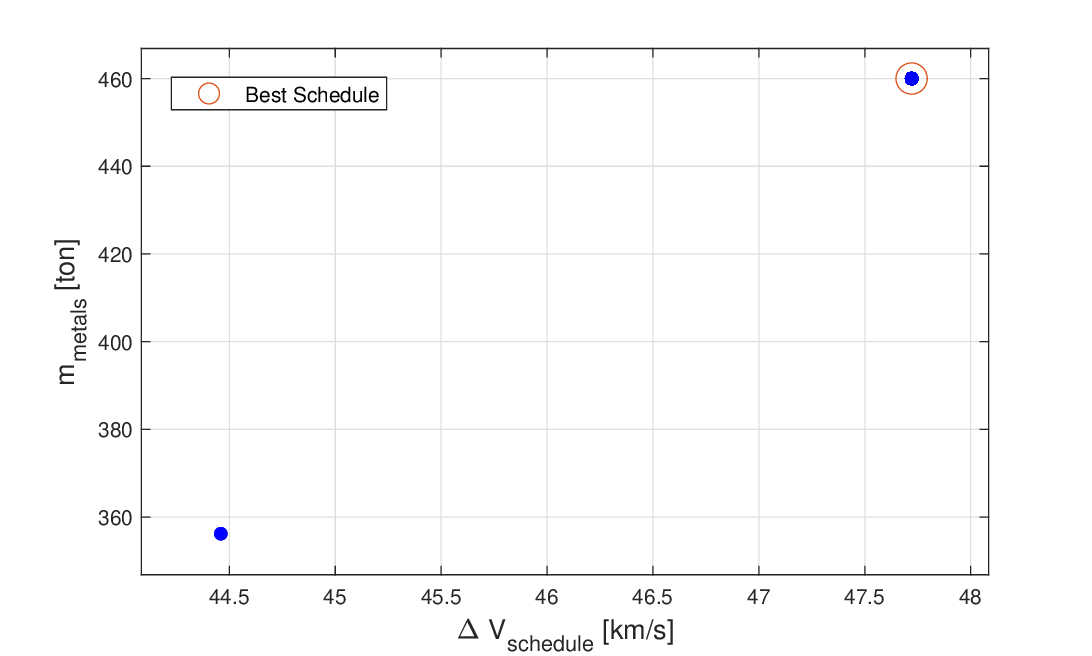}
  \caption{Pareto front with maximum mining rate if 4 metallic asteroids are visited.}
  \label{fig:p_800_0_xxxx}
   \end{center}
\end{figure}

\begin{figure}[!htb]
 \begin{center}
  \includegraphics[width=0.8\textwidth]{ 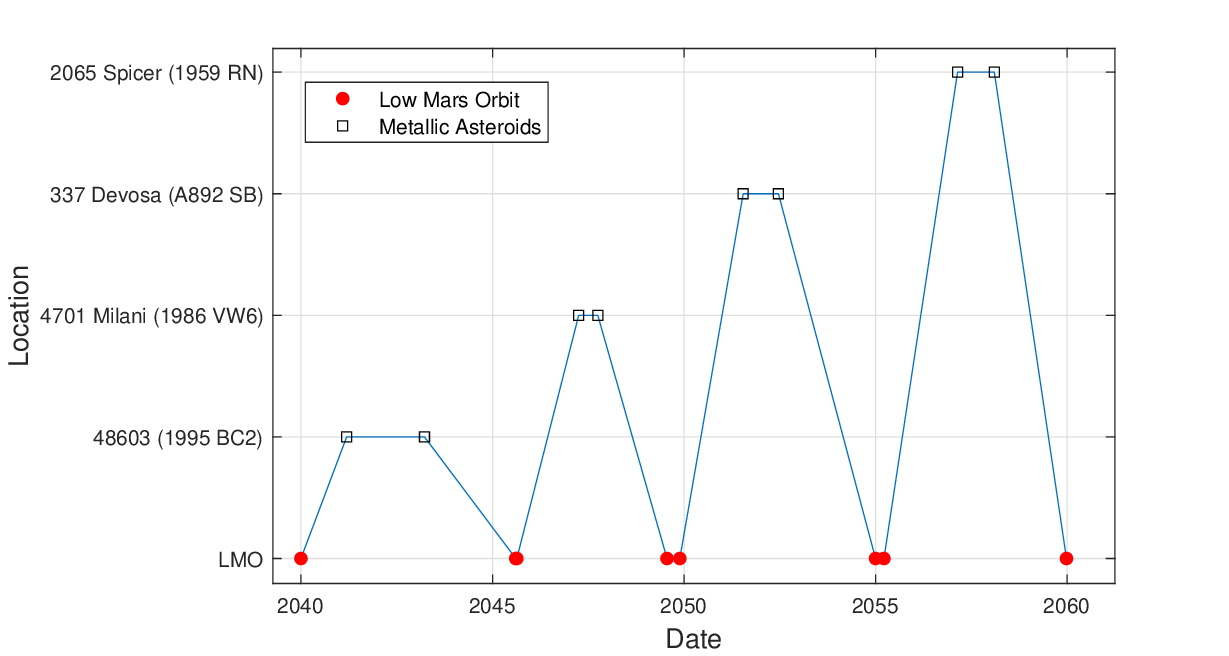}
  \caption{Best schedule with maximum mining rate if only metallic asteroids are visited.}
  \label{fig:s_800_0_xxxx}
   \end{center}
\end{figure}

In Fig.~\ref{fig:p_800_0_xxxx} and Fig.~\ref{fig:s_800_0_xxxx} the outcome of the problem solved considering the maximum mining rate is shown. 

\begin{figure}[!htb]
 \begin{center}
  \includegraphics[width=0.8\textwidth]{ 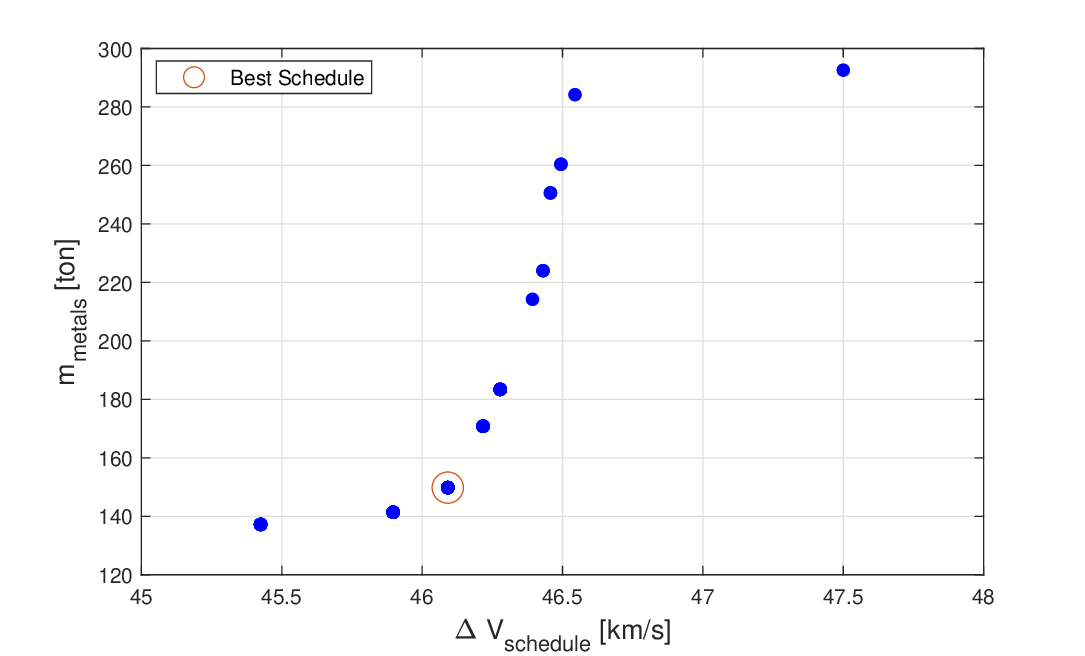}
  \caption{Pareto front with minimum mining rate if metallic asteroids are visited with the lower bound of \SI{10}{ton} on the mass delivered for each trip.}
  \label{fig:p_100_10_xxxx}
   \end{center}
\end{figure}

\begin{figure}[!htb]
 \begin{center}
  \includegraphics[width=0.8\textwidth]{ 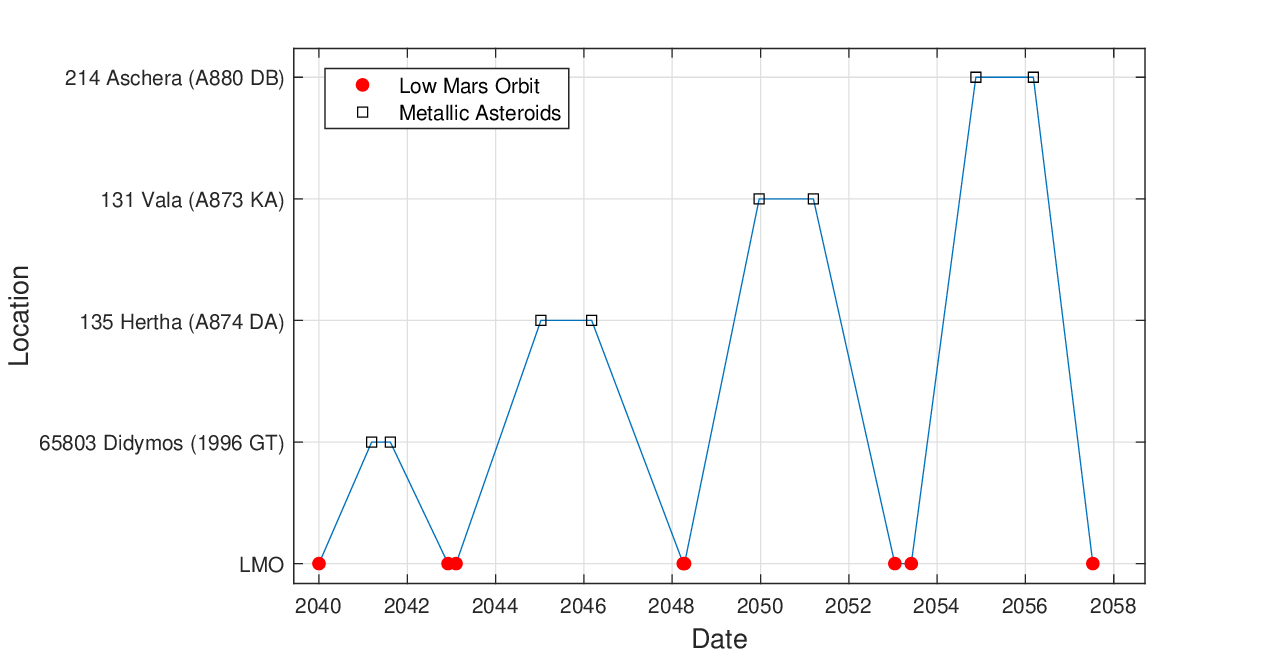}
  \caption{Best schedule with minimum mining rate if metallic asteroids are visited with the lower bound of \SI{10}{ton} on the mass delivered for each trip.}
  \label{fig:s_100_10_xxxx}
   \end{center}
\end{figure}

With the addition of the constraints provided in Table~\ref{tab:limit_10_80}, for each return transfer on LMO, a minimum mass of material to be delivered is imposed. The optimization problems are executed with these constraints and the results are shown in Figures~\ref{fig:p_100_10_xxxx}-\ref{fig:s_100_10_xxxx}.

\subsection{Metallic-Carbonaceous Asteroids in the schedule}
Considering the cargo spacecraft (Table~\ref{tab:Starship}), the propellant capability guar\-an\-tees a $\Delta V$ of \SI{6.4}{km/s}. A schedule under this condition allows to visit 2 pairs of asteroids in 20 years with a single spacecraft. It is thus possible to mine on two metallic asteroids and refuel on the two associated carbonaceous asteroids. 
\begin{figure}[!htb]
 \begin{center}
  \includegraphics[width=0.75\textwidth]{ 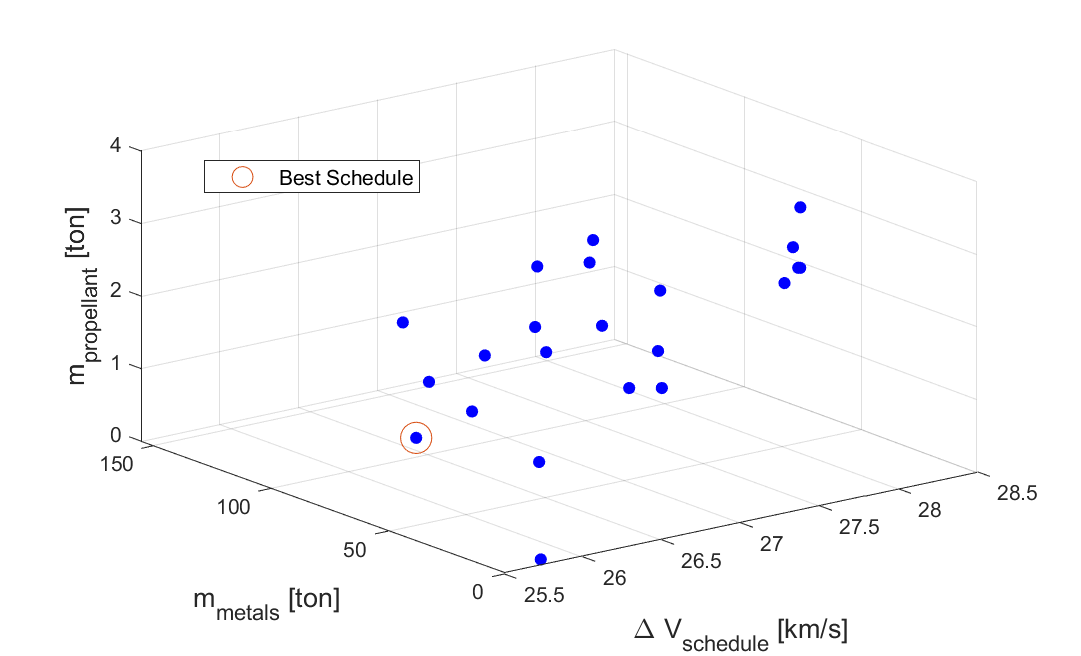}
  \caption{Pareto front, minimum mining rate, metallic and carbonaceous asteroids are visited.}
  \label{fig:p_100_0_xcxc}
   \end{center}
\end{figure}
\begin{figure}[!htb]
 \begin{center}
  \includegraphics[width=0.8\textwidth]{ 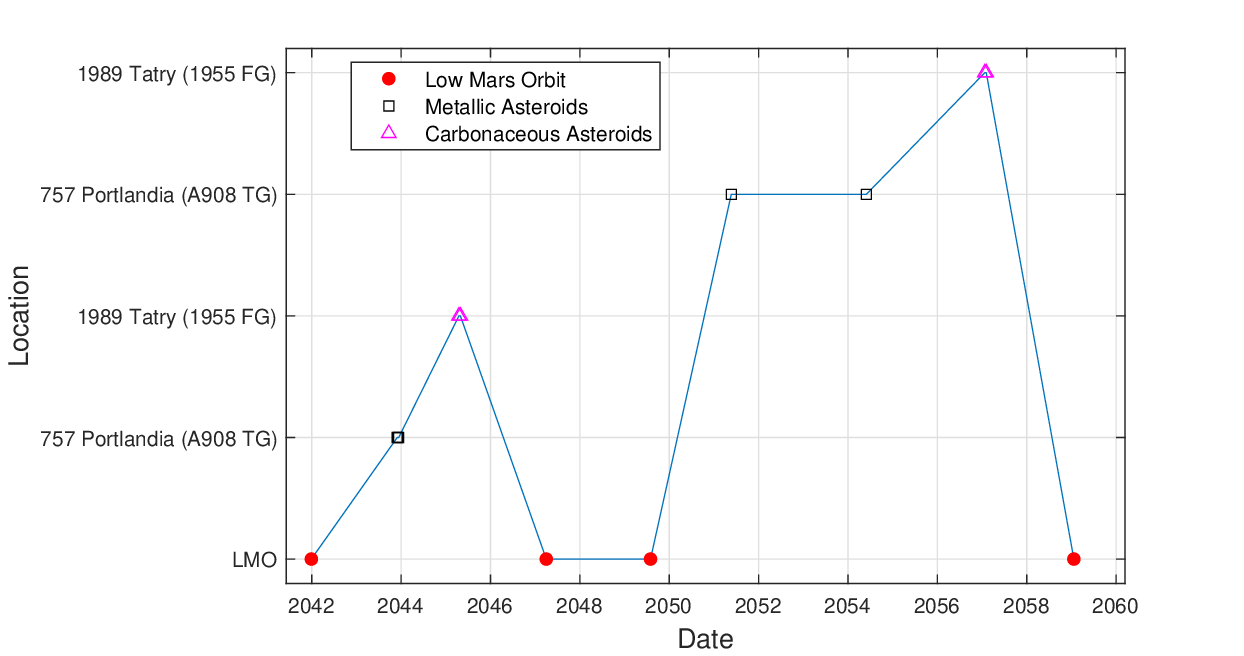}
  \caption{Best schedule, minimum mining rate, metallic and carbonaceous asteroids are visited.}
  \label{fig:s_100_0_xcxc}
   \end{center}
\end{figure}

Fig.~\ref{fig:p_100_0_xcxc} shows the Pareto front coming from the optimization performed with the minimum value of the mining rate. The best solution is circled. The best schedule executed with one single spacecraft with the minimum mining rate of \SI{100}{kg/day} \cite{dorrington_rate} is shown in Fig.~\ref{fig:s_100_0_xcxc}. Note that the plan consists of visiting a total of four asteroids in which the first and the third ones are metallic, while the second and the fourth ones are carbonaceous.

The mass delivered on Mars could be used for additive manufacturing of rovers and for some repairs. For example, we take into consideration the Perseverance rover \citep{perseverance} whose mass is equal to \SI{1025}{kg}. It is estimated, for each schedule, how many rovers of this mass can be built. The available mass varies based on the extraction rate and the schedule chosen among the possible options given by the multi-objective optimization. If a large amount of $m_{metals}$ is available, the metals could be used for the additive construction of houses that could provide an additional benefit in terms of protection from radiation and thus ensure the growth of the colony. Starting from the data provided by \cite{kading2015317}, the volume of material required for the construction of 12 habitats is estimated to be \SI{370}{m^3}. Estimating the density of metallic meteorites to be about \SI{4000}{kg/m^3} \citep{meteorite}, a mass of about \SI{1480}{ton} of metals extracted is necessary to build 12 habitats. Each habitat has a base diameter of \SI{8.40}{m} and a thickness of the dome of \SI{0.25}{m}. With a volume per-dome of \SI{128}{m^3}, the 12-dome base could approximately support 15 individuals \citep{kading2015317}. Note that the actual number depends on the base configuration and the laboratories needs.

In the case presented in Fig.~\ref{fig:s_100_0_xcxc} we obtain $m_{metals} = \SI{111.6}{ton}$ with a $\Delta V = \SI{26.59}{km/s}$, so about 100 rovers with a mass equal to that of Perseverance can be built. 

\begin{figure}[!htb]
 \begin{center}
  \includegraphics[width=0.8\textwidth]{ 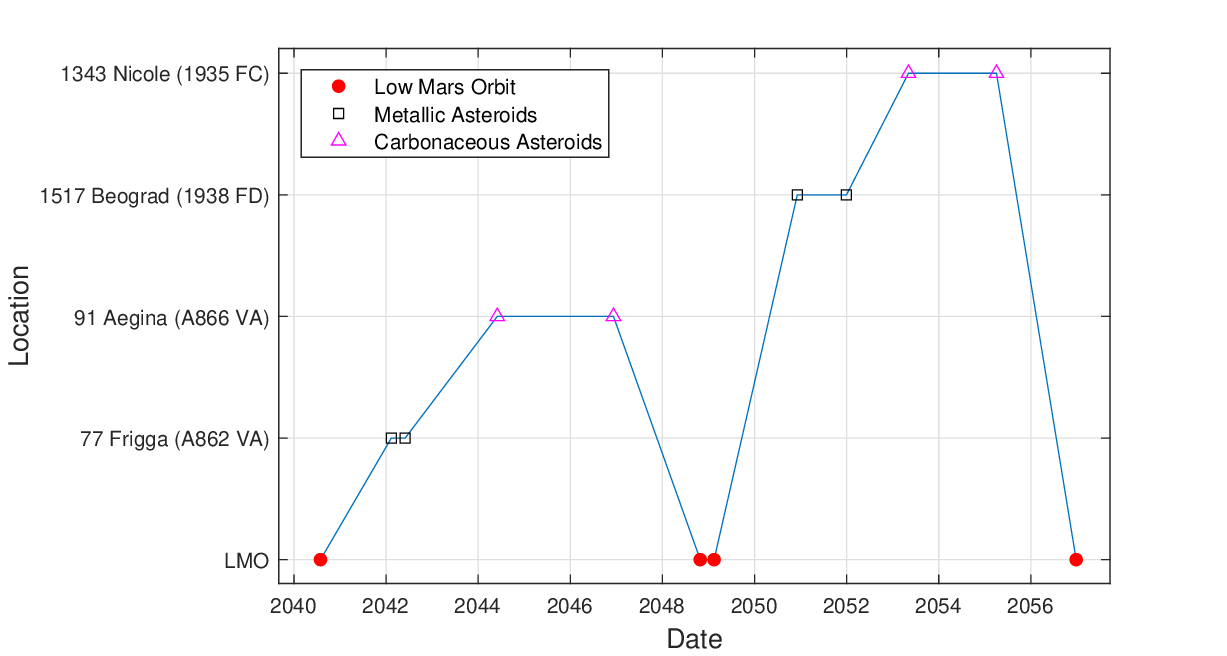}
  \caption{Best schedule with maximum mining rate if metallic and carbonaceous asteroids are visited.}
  \label{fig:s_800_0_xcxc}
   \end{center}
\end{figure}

If we consider the maximum mining rate of \SI{800}{kg/day}, we get the schedule in Fig.~\ref{fig:s_800_0_xcxc}. This schedule allows the delivery of $m_{metals}= \SI{203}{ton}$ with a $\Delta V = \SI{28.12}{km/s}$ and so the construction of 1 housing module and about 70 rovers in 20 years is possible.

\begin{figure}[!htb]
 \begin{center}
  \includegraphics[width=0.8\textwidth]{ 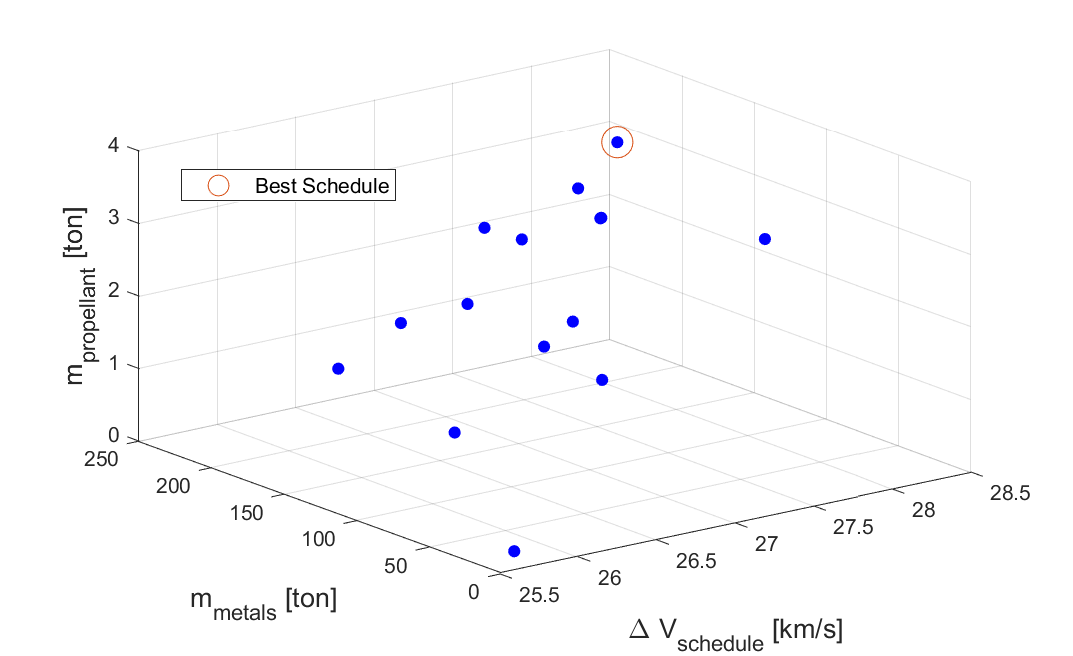}
  \caption{Pareto front with maximum mining rate if metallic and carbonaceous asteroids are visited.}
  \label{fig:p_800_0_xcxc}
   \end{center}
\end{figure}

One could think of the possibility of using more spacecraft that can guarantee a greater mass delivery. According to the multi-objective optimization one spacecraft is associated to each optimal solutions of the Pareto front. By considering the Pareto front in Fig.~\ref{fig:p_800_0_xcxc} and using one spacecraft for each of the elements of the Pareto front, that are $70$, it is possible to deliver \SI{12095}{ton} of metallic material to LMO. It is sufficient for building of 8 housing complexes, of 12 habitats each, that can host approximately $120$ people. Moreover, there is enough material to guarantee the repairs needed and the assembly of more than 500 rovers. 

With the addition of the constraint given in Table~\ref{tab:limit_10_80}, for each return transfer on LMO, a minimum mass of material to be delivered is imposed. The opti\-mi\-za\-tion problems are executed with these constraints and the results are here shown in the following figures (Fig.~\ref{fig:p_100_10_xcxc}-\ref{fig:s_800_80_xcxc}).

\begin{figure}[!htb]
 \begin{center}
  \includegraphics[width=0.8\textwidth]{ 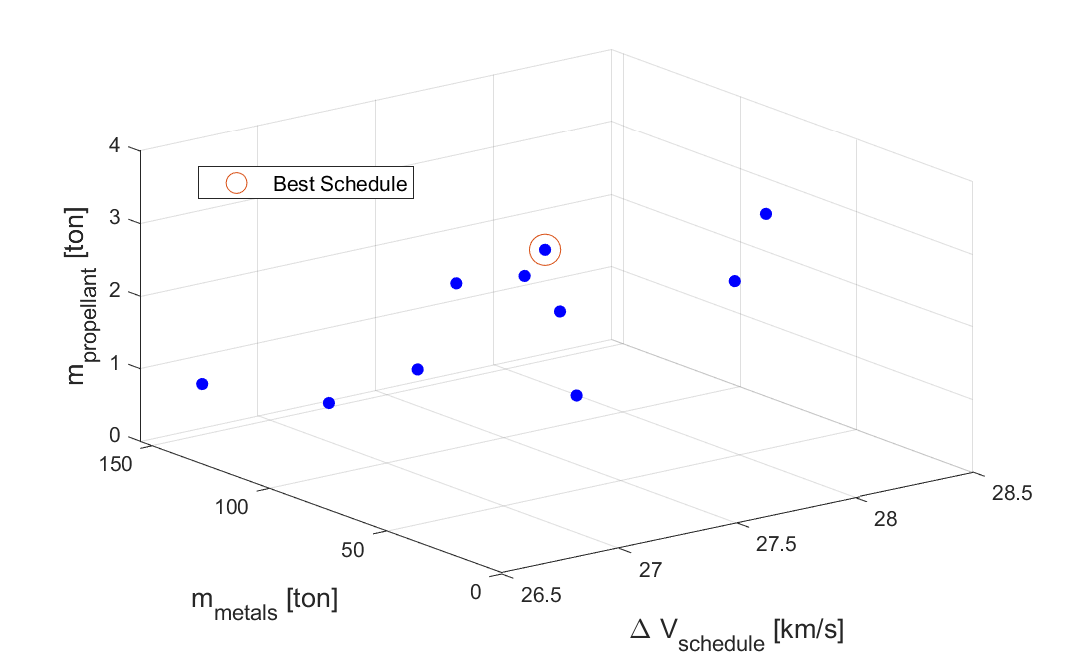}
  \caption{Pareto front with minimum mining rate if metallic and carbonaceous asteroids are visited with the lower bound of \SI{10}{ton} on the mass delivered for each trip.}
  \label{fig:p_100_10_xcxc}
   \end{center}
\end{figure}

\begin{figure}[!htb]
 \begin{center}
  \includegraphics[width=0.8\textwidth]{ 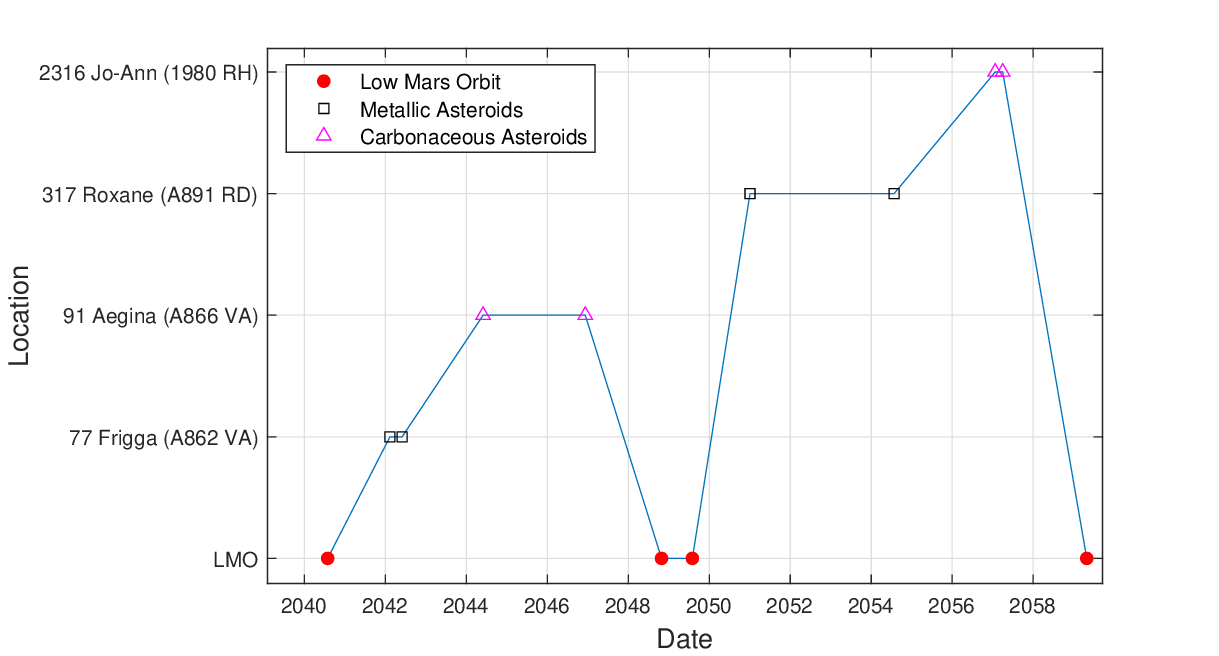}
  \caption{Best schedule with minimum mining rate if metallic and carbonaceous asteroids are visited with the lower bound of \SI{10}{ton} on the mass delivered for each trip.}
  \label{fig:s_100_10_xcxc}
   \end{center}
\end{figure}

\begin{figure}[!htb]
 \begin{center}
  \includegraphics[width=0.8\textwidth]{ 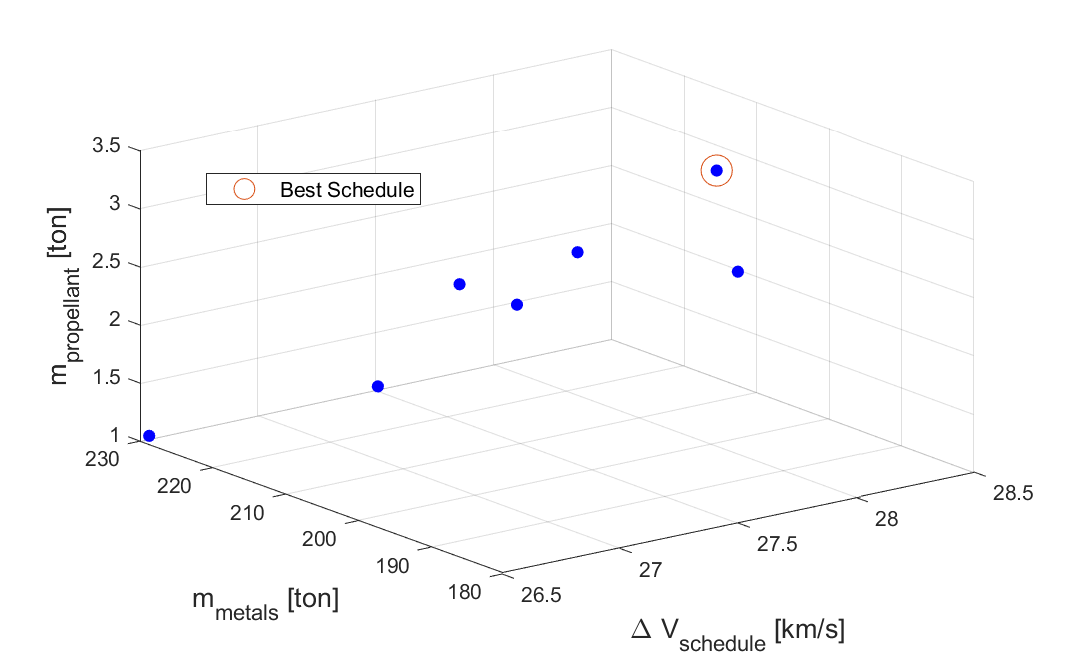}
  \caption{Pareto front with maximum mining rate if metallic and carbonaceous asteroids are visited with the lower bound of \SI{80}{ton} on the mass delivered for each trip.}
  \label{fig:p_800_80_xcxc}
   \end{center}
\end{figure}

\begin{figure}[!htb]
 \begin{center}
  \includegraphics[width=0.8\textwidth]{ 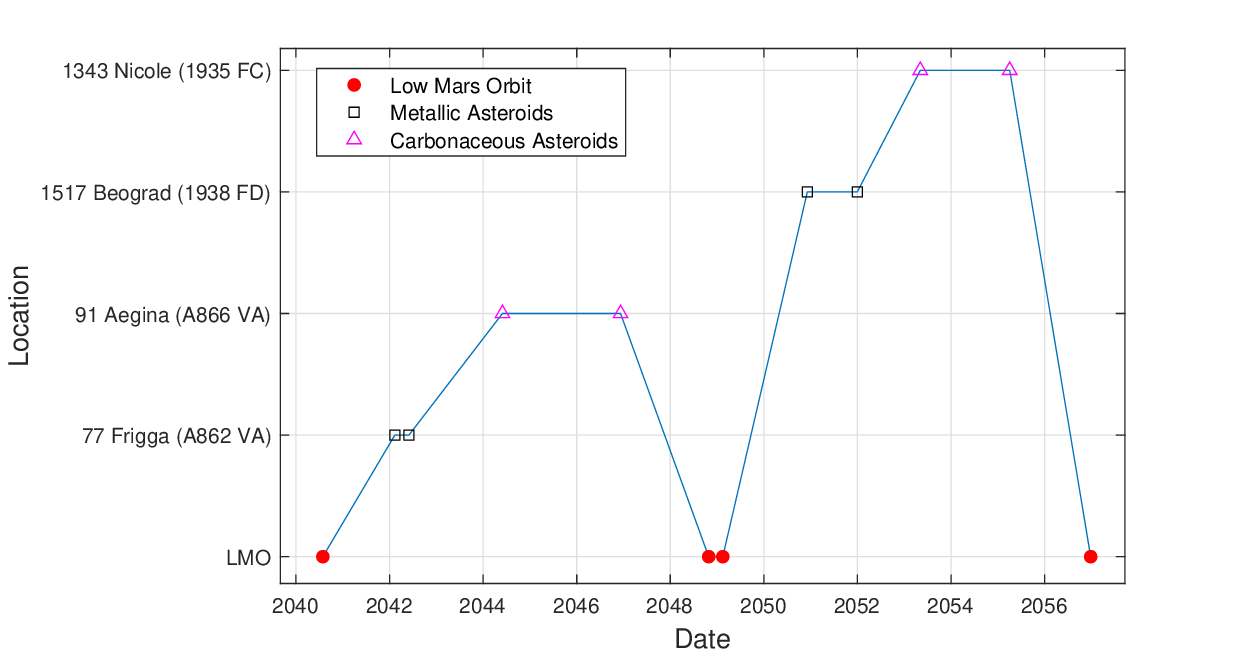}
  \caption{Best schedule with maximum mining rate if metallic and carbonaceous asteroids with the lower bound of \SI{80}{ton} on the mass delivered for each trip.}
  \label{fig:s_800_80_xcxc}
   \end{center}
\end{figure}

\subsection{Asteroids and Depot in the schedule}\label{sec:depot_schedule}
The schedules in which the depot is taken into account as an additional source are computed. 

First we analyzed the case in which the spacecraft visits only the depot and the metallic asteroids. The best schedules outgoing from the optimizations conducted for the several cases are reported below.

\begin{figure}[!htb]
\centering
 \begin{center}
  \includegraphics[width=0.8\textwidth]{ 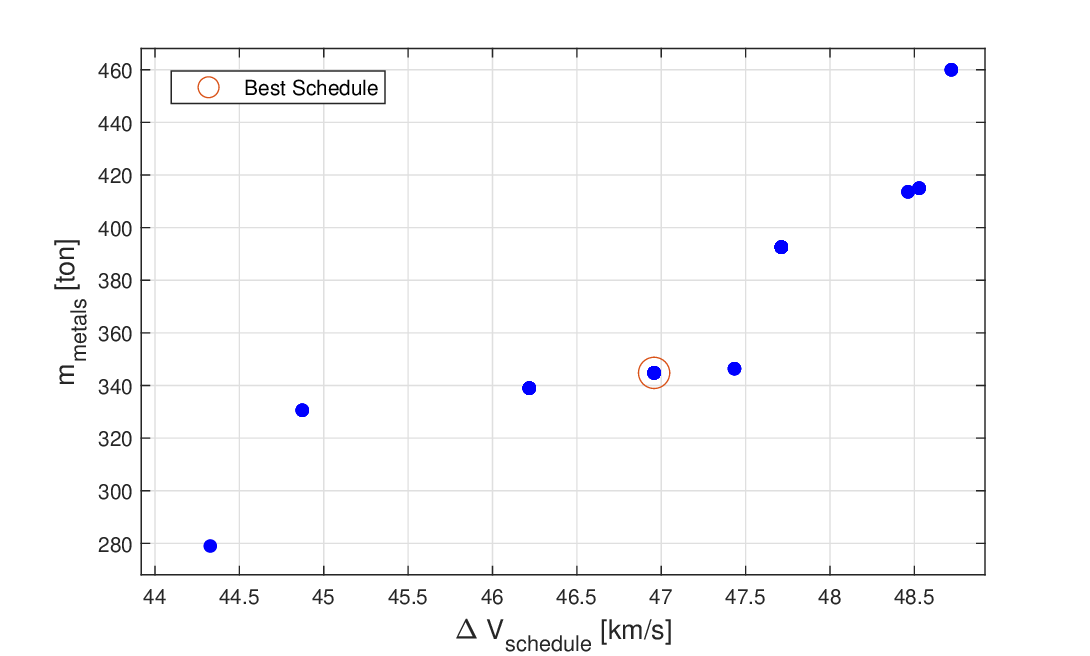}
  \caption{Pareto front with minimum mining rate if depot and metallic asteroids are visited with no lower bound on the mass delivered.}
  \label{fig:xL2_100_0_pf}
   \end{center}
\end{figure}

Imposing the lower bound on the mass delivered to LMO for each trip (shown in Table \ref{tab:limit_10_80}) the following results are obtained.

Finally, by adding the depot in the optimization in which metallic and carbonaceous asteroids are visited, the best schedule allows to use only the depot as a source, since it can guarantee the maximum transportable mass of metals and propellant. The results are the same both with the minimum and maximum mining rate. This schedule allows to deliver $m_{metals}=\SI{345}{ton}$ to LMO with one spacecraft and a $\Delta V_{schedule}=\SI{36.58}{km/s}$.

\subsection{Comparison of the optimized schedules}

\begin{figure}[!htb]
\centering
 \begin{center}
  \includegraphics[width=0.8\textwidth]{ 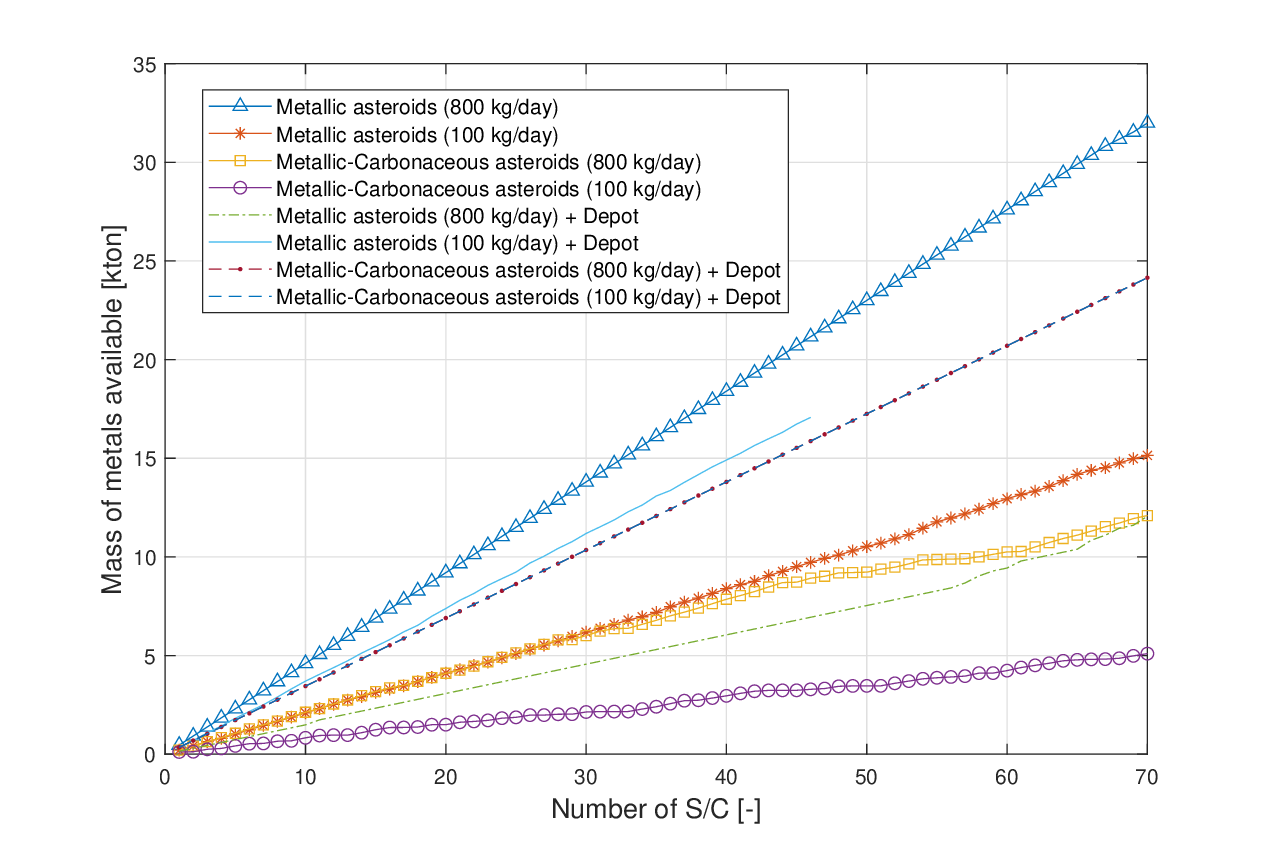}
  \caption{Mass of metals available depending on the number of spacecraft used.}
  \label{fig:comparison_no_limits}
   \end{center}
\end{figure}

\begin{figure}[!htb]
\centering
 \begin{center}
  \includegraphics[width=0.8\textwidth]{ 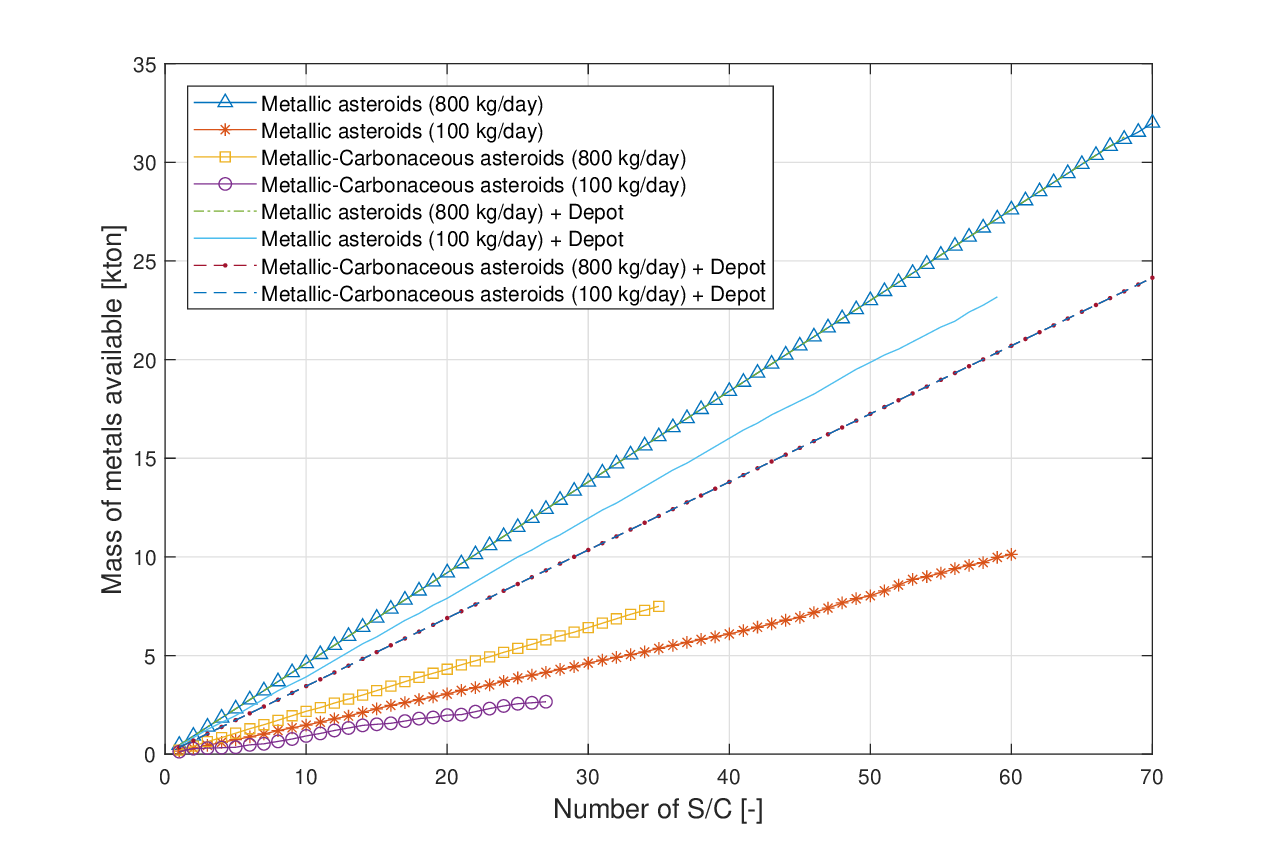}
  \caption{Mass of metals available depending on the number of spacecraft used by imposing the lower bounds on the $m_{metals}$.}
  \label{fig:comparison_yes_limits}
   \end{center}
\end{figure}

To conclude, we present a comparison between  the optimized cases. 
Fig.~\ref{fig:comparison_no_limits} and \ref{fig:comparison_yes_limits} show how the available mass increases with the number of spacecraft employed. The cases taken into account are 'Metallic Asteroids', 'Metallic-Carbonaceous Asteroids', 'Metallic Asteroids + Depot', 'Metallic-Car\-bo\-na\-ceous Asteroids + Depot'. Fig.~\ref{fig:comparison_no_limits} is the outcome of the optimizations with no lower bounds imposed on $m_{mined}$. Fig.~\ref{fig:comparison_yes_limits} is the outcome of the optimizations with the imposition of lower bounds on $m_{mined}$.

Notice that some lines in Fig.~\ref{fig:comparison_yes_limits} are shorter with respect to the ones in Fig.~\ref{fig:comparison_no_limits} because of the lower number of elements in the Pareto front. Comparing the two figures one can see how the imposed limitations result in higher in\-cli\-na\-tions for some of the lines in Fig.~\ref{fig:comparison_yes_limits}. This is the result of a better effectiveness of the schedules, i.e. in the case in which only metallic asteroids are considered with a mining rate of \SI{100}{kg/day}, with 20 spacecrafts it is possible to deliver almost  \SI{5000}{ton} in
\ref{fig:comparison_yes_limits}, but in Fig.~\ref{fig:comparison_no_limits} the mass of metals corresponding to 20 spacecraft is smaller. 


\section{Discussion}

The results given in the previous section indicate that asteroid-mining cam\-paigns to provide the Mars base with metals are feasible from the logistics point of view, provided a certain technological level of the spacecraft involved and in-situ production of fuel on Mars. One of the most interesting and encouraging outcomes of the study is that asteroid mining can be more advantageous than direct transfer of materials from Earth. This is clearly shown by the optimal schedules, where the Sun-Earth L2 depot could be chosen along with the asteroids (Section~\ref{sec:depot_schedule}), however, along with L2, trips to asteroids are still found in the schedules. Of course, the Earth will still remain an indispensable supplier of all kinds of necessities for the Martian base, but given the option of delivering heavier and bulky metals from asteroids, the cargo spacecraft from Earth can have more room for other commodities, which can be produced only on the Earth.

The available mass of metals depends on the mining rate and the stay time on the asteroids. The metals can be employed in the additive manufacturing \citep{Owens2015} and additive construction \citep{kading2015317}, as well as for repairs.
The feasibility of this mission strongly depends on the technologies that allow in-situ propellant production. The convenience of this mission depends, instead, on the rate of extraction of metals, which is influenced by the power available and, in turn, influences the mass of metals to be transferred to Mars. 
The mining rate can be increased by increasing the power available and so by increasing the dimensions of the solar arrays of the mining equipment \citep{dorrington_rate}. Instead, for what regards the propellant production rate, the current technologies cannot guarantee a sufficient rate \citep{zubrin_propellant}. Having a stay time of, at most, hundreds of days on the carbonaceous asteroids, a propellant production rate of hundreds of kg/day is necessary. Such a production rate has not yet been achieved with current technologies \citep{zubrin_propellant}. So far, a production rate of \SI{2}{kg/day} \citep{zubrin2019} is considered in the studies related to the first manned mission on Mars.

\section{Conclusion}

From a logistics and organizational point of view, this mission is feasible and can guarantee a certain independence of the colony from Earth in terms of materials for additive manufacturing and additive construction. However, the results strongly depend on the mining rate and propellant production rate. Given the technological limitations surrounding the production of propellant on Mars and on the carbonaceous asteroids, further research and improvements in that field will be necessary to carry out the mission. Alternative propulsion methods such as solar sails or nuclear propulsion could also be studied.

To conclude, it can be said that by improving the rate of production of the propellant, an optimized supply chain is obtained which leads to the growth of the colony thanks to the possibility of building habitats with the mined metals on asteroids. In this way the radiation protection can be ensured. Therefore, the purpose of the study was achieved. Several plans are evaluated by varying some parameters. They all make it possible to exploit the extraction of metals from asteroids by minimizing the variation in total velocity and maximizing the mass deliverable to the colony. The available mass of metallic material largely satisfies the requirements of a colony of varying size, depending on the number of spacecraft used. 

\newpage

\section*{Appendix A: Pairs of Metallic-Carbonaceous Asteroids} \label{appendixA}

\begin{table}[!htb]
\centering
\begin{tabular}{ll}
\multicolumn{1}{l}{Metallic Asteroids} & \multicolumn{1}{l}{Carbonaceous Asteroids} \\ \hline
44 Nysa (A857 KA) & 1076 Viola (1926 TE) \\ \hline
77 Frigga (A862 VA) & 91 Aegina (A866 VA) \\ \hline
214 Aschera (A880 DB) & 4124 Herriot (1986 SE) \\ \hline
261 Prymno (A886 UA) & 3192 A'Hearn (1982 BY1) \\ \hline
272 Antonia (A888 CA) & 257 Silesia (A886 GB) \\ \hline
317 Roxane (A891 RD) & 2316 Jo-Ann (1980 RH) \\ \hline
332 Siri (A892 FH) & 2852 Declercq (1981 QU2) \\ \hline
338 Budrosa (A892 SF) & 4944 Kozlovskij (1987 RP3) \\ \hline
757 Portlandia (A908 TG) & 1989 Tatry (1955 FG) \\ \hline
1039 Sonneberga (A924 WP) & 743 Eugenisis (A913 DD) \\ \hline
1046 Edwin (A924 XF) & 195 Eurykleia (A879 HA) \\ \hline
1201 Strenua (1931 RK) & 743 Eugenisis (A913 DD) \\ \hline
1352 Wawel (1935 CE) & 1766 Slipher (1962 RF) \\ \hline
1517 Beograd (1938 FD) & 1343 Nicole (1935 FC) \\ \hline
1541 Estonia (1939 CK) & 195 Eurykleia (A879 HA) \\ \hline
2306 Bauschinger (1939 PM) & 3645 Fabini (1981 QZ) \\ \hline
2560 Siegma (1932 CW) & 58 Concordia (A860 FA) \\ \hline
3575 Anyuta (1984 DU2) & 7405 (1988 FF) \\ \hline
4342 Freud (1987 QO9) & 301 Bavaria (A890 WA) \\ \hline
4548 Wielen (2538 P-L) & 1989 Tatry (1955 FG) \\ \hline
4701 Milani (1986 VW6) & 8008 (1988 TQ4) \\ \hline
5576 Albanese (1986 UM1) & 127 Johanna (A872 VB) \\ \hline
\title{Pairs of Metallic-Carbonaceous Asteroids}
\label{tab:pairs}
\end{tabular}
\end{table}
\clearpage

 \section*{Appendix B: Optimization setup\footnote{@gamultiobj by MATLAB is based on non-integer programming}}\label{appendixB}

\begin{table}[!htb]
\centering
\begin{tabular}{cl}
\hline
Case & \multicolumn{1}{c}{Input } \\ \hline
4 Metallic Asteroids 
& \begin{tabular}[c]{@{}l@{}}- Number of Variables: 4 \\ - Lower Bound: {[}0.5 0.5 0.5 0.5{]} \\ - Upper Bound: {[}122.49 122.49 122.49 122.49{]}  \end{tabular}  \\ \hline
2 Pairs of Metallic-Carbonaceous Asteroids 
& \begin{tabular}[c]{@{}l@{}}- Number of Variables: 2\\ - Lower Bound: {[}0.5 0.5{]}\\ - Upper Bound: {[}22.49 22.49{]}   \end{tabular} \\ \hline
\end{tabular}
\label{tab:setup1}
\end{table}


\begin{table}[!htb]
\centering
\begin{tabular}{cl}
\hline
Case & \multicolumn{1}{c}{Input } \\ \hline
Metallic Asteroids + Depot  & \begin{tabular}[c]{@{}l@{}}- Number of Variables: 4 \\ - Lower Bound: {[}0.5 0.5 0.5 0.5{]} \\ - Upper Bound: {[}123.49 123.49 123.49 123.49{]}  \end{tabular}  \\ \hline
Metallic-Carbonaceous Asteroids  + Depot & \begin{tabular}[c]{@{}l@{}}- Number of Variables: 3\\ - Lower Bound: {[}0.5 0.5 0.5{]}\\ - Upper Bound: {[}23.49 23.49 23.49{]}   \end{tabular} \\ \hline
\end{tabular}
\label{tab:setup2}
\end{table}
\clearpage

 \section*{Appendix C: Optimization Options} \label{appendixC}

\begin{table}[!htb]
\centering
\label{tab:options}
\centering
\begin{tabular}{l l}
\hline
Population & \begin{tabular}[c]{@{}l@{}}- Population size: 200\\ - Initial range: {[}-10,10{]} \\ - Creation Function: Uniform\end{tabular} \\ \hline
Selection & \begin{tabular}[c]{@{}l@{}}- Selection Function: Tournament\\ - Tournament size: 2\end{tabular} \\ \hline
Reproduction & - Crossover Fraction: 0.8 \\ \hline
Mutation & \begin{tabular}[c]{@{}l@{}}- Mutation Function: Uniform \\ - Rate: 0.1 \end{tabular} \\ \hline
Crossover & \begin{tabular}[c]{@{}l@{}}- Crossover function: Two point\end{tabular} \\ \hline
Migration & \begin{tabular}[c]{@{}l@{}}- Direction: Forward\\ - Fraction: 0.2\\ - Interval: 20\end{tabular} \\ \hline
Multiobjective Problem Setting & \begin{tabular}[c]{@{}l@{}}- Distance Measure Function: @distancecrowding\\ - Pareto Front Population Fraction: 0.35\end{tabular} \\ \hline
Hybrid Function & - Hybrid Function: None \\ \hline
Stopping Criteria & \begin{tabular}[c]{@{}l@{}}- Generations: $100 \times number \ of \ variables$\\ - Time limit: Inf\\ - Fitness limit: -Inf\\ - Stall Generations: 100\\ - Stall Time Limit: Inf\\ - Function Tolerance: 1e-4\\ - Constraint tolerance: 1e-3\end{tabular} \\ \hline
\end{tabular}
\end{table}
\clearpage

 \section*{Appendix D: Results with no lower bound on the mass delivered and one spacecraft} \label{appendixD}

\begin{table}[!htb]
\centering
\label{tab:results_no_limits}
\centering
\begin{tabular}{ccc}
\hline 
Optimization Case & $\Delta V_{schedule} \ [km/s]$ & $m_{metals} \ [ton]$ \\ \hline
\begin{tabular}[c]{@{}c@{}}\textbf{Metallic Asteroids}\\ $\dot m_{mining} = \SI{100}{kg/day}$\end{tabular} & 44.98 & 204.4 \\ \hline
\begin{tabular}[c]{@{}c@{}}\textbf{Metallic Asteroids}\\ $\dot m_{mining} = \SI{800}{kg/day}$\end{tabular} & 47.72 & 460 \\ \hline
\begin{tabular}[c]{@{}c@{}}\textbf{Metallic-Carbonaceous Asteroids}\\ $\dot m_{mining} = \SI{100}{kg/day}$\end{tabular} & 26.59 & 111.6 \\ \hline
\begin{tabular}[c]{@{}c@{}}\textbf{Metallic-Carbonaceous Asteroids}\\ $\dot m_{mining} = \SI{800}{kg/day}$\end{tabular} & 28.12 & 203 \\ \hline
\begin{tabular}[c]{@{}c@{}}\textbf{Metallic Asteroids + Depot}\\ $\dot m_{mining} = \SI{100}{kg/day}$\end{tabular} & 46.96 & 344.8 \\ \hline
\begin{tabular}[c]{@{}c@{}}\textbf{Metallic Asteroids + Depot }\\ $\dot m_{mining} = \SI{800}{kg/day}$\end{tabular} & 43.87 & 148.6 \\ \hline
\begin{tabular}[c]{@{}c@{}}\textbf{Metallic-Carbonaceous Asteroids + Depot}\\ $\dot m_{mining} = \SI{100}{kg/day}$\end{tabular} & 36.58 & 345 \\ \hline
\begin{tabular}[c]{@{}c@{}}\textbf{Metallic-Carbonaceous Asteroids + Depot}\\ $\dot m_{mining} = \SI{800}{kg/day}$\end{tabular} & 36.58 & 345 \\ \hline
\end{tabular}
\end{table}
\clearpage

 \section*{Appendix E: Results with lower bounds on the mass delivered and one spacecraft} \label{appendixE}

\begin{table}[!htb]
\centering
\label{tab:results_yes_limits}
\centering
\begin{tabular}{ccc}
\hline
Optimization Case & $\Delta V_{schedule} \ [km/s]$ & $m_{metals} \ [ton]$ \\ \hline
\begin{tabular}[c]{@{}c@{}} \textbf{Metallic Asteroids}  \\ $\dot m_{mining} = \SI{100}{kg/day}$\\$m_{metals}^{min} = \SI{10}{kg}$\end{tabular} & 46.09 & 149.8 \\ \hline
\begin{tabular}[c]{@{}c@{}}\textbf{Metallic Asteroids}\\ $\dot m_{mining} = \SI{800}{kg/day}$\\$m_{metals}^{min} = \SI{80}{kg}$\end{tabular} & 46.8 & 460 \\ \hline
\begin{tabular}[c]{@{}c@{}}\textbf{Metallic-Carbonaceous Asteroids}\\ $\dot m_{mining} = \SI{100}{kg/day}$\\$m_{metals}^{min} = \SI{10}{kg}$\end{tabular} & 27.93 & 126 \\ \hline
\begin{tabular}[c]{@{}c@{}}\textbf{Metallic-Carbonaceous Asteroids}\\ $\dot m_{mining} = \SI{800}{kg/day}$\\$m_{metals}^{min} = \SI{80}{kg}$\end{tabular} & 28.12 & 203 \\ \hline
\begin{tabular}[c]{@{}c@{}}\textbf{Metallic Asteroids + Depot}\\ $\dot m_{mining} = \SI{100}{kg/day}$\\$m_{metals}^{min} = \SI{10}{kg}$ \end{tabular} & 47.32 & 419.2 \\ \hline
\begin{tabular}[c]{@{}c@{}}\textbf{Metallic Asteroids + Depot} \\ $\dot m_{mining} = \SI{800}{kg/day}$\\$m_{metals}^{min} = \SI{80}{kg}$\end{tabular} & 46.34 & 460 \\ \hline
\begin{tabular}[c]{@{}c@{}}\textbf{Metallic-Carbonaceous Asteroids + Depot}\\ $\dot m_{mining} = \SI{100}{kg/day}$\\$m_{metals}^{min} = \SI{10}{kg}$\end{tabular} & 36.58 & 345 \\ \hline
\begin{tabular}[c]{@{}c@{}}\textbf{Metallic-Carbonaceous Asteroids + Depot}\\ $\dot m_{mining} = \SI{800}{kg/day}$\\$m_{metals}^{min} = \SI{80}{kg}$\end{tabular} & 36.58 & 345 \\ \hline
\end{tabular}
\end{table}

\end{document}